\documentclass[pdflatex,sn-nature]{sn-jnl}

\usepackage{graphicx}%
\usepackage{multirow}%
\usepackage{amsmath,amssymb,amsfonts}%
\usepackage{amsthm}%
\usepackage{mathrsfs}%
\usepackage[title]{appendix}%
\usepackage{xcolor}%
\usepackage{textcomp}%
\usepackage{manyfoot}%
\usepackage{booktabs}%
\usepackage{algorithm}%
\usepackage{algorithmicx}%
\usepackage{algpseudocode}%
\usepackage{listings}%
\usepackage{rotating}

\theoremstyle{thmstyleone}%
\theoremstyle{thmstyletwo}%

\theoremstyle{thmstylethree}%

\begin{document}

\title[GRB 220706A: a gamma-ray burst with a month-long engine and a luminous supernova]{GRB 220706A: a gamma-ray burst with a month-long engine and a luminous supernova}


\author*[1,2]{\fnm{Benjamin P.} \sur{Gompertz}}\email{b.gompertz@bham.ac.uk}
\author[3]{\fnm{Nusrin} \sur{Habeeb}}
\author[4]{\fnm{Dheeraj R.} \sur{Pasham}}
\author[5]{\fnm{Antonio} \sur{de Ugarte Postigo}}
\author[6,7]{\fnm{Daniele B.} \sur{Malesani}}
\author[3]{\fnm{Phil A.} \sur{Evans}}
\author[3]{\fnm{Kim L.} \sur{Page}}
\author[3]{\fnm{Ben} \sur{Rayson}}
\author[8]{\fnm{J. Feliciano} \sur{Agüí Fernández}}
\author[9]{\fnm{Michael} \sur{Bremer}}
\author[10,11]{\fnm{Andrew J.} \sur{Levan}}
\author[12]{\fnm{Genevieve} \sur{Schroeder}}
\author[3]{\fnm{Nial R.} \sur{Tanvir}}
\author[13]{\fnm{Christina C.} \sur{Thöne}}
\author[14]{\fnm{Eric} \sur{Burns}}
\author[15]{\fnm{Valerio} \sur{D'Elia}}
\author[16]{\fnm{Massimiliano} \sur{De Pasquale}}
\author[1,2]{\fnm{} \sur{Dimple}}
\author[17]{\fnm{Dieter} \sur{Hartmann}}
\author[18]{\fnm{Pall} \sur{Jakobsson}}
\author[19]{\fnm{Sylvio} \sur{Klose}}
\author[20]{\fnm{Antonio} \sur{Martin-Carrillo}}
\author[21]{\fnm{Matt} \sur{Nicholl}}
\author[19]{\fnm{Ana M.} \sur{Nicuesa Guelbenzu}}
\author[1,2]{\fnm{David} \sur{O'Neill}}
\author[22]{\fnm{Giovanna} \sur{Pugliese}}
\author[23]{\fnm{Arne} \sur{Rau}}
\author[24]{\fnm{Andrea} \sur{Rossi}}
\author[25]{\fnm{Jesper} \sur{Sollerman}}
\author[3]{\fnm{Rhaana} \sur{Starling}}
\author[6,7]{\fnm{Darach} \sur{Watson}}
\author[1,2]{\fnm{Isabelle} \sur{Worssam}}
\author[1,2]{\fnm{Makenzie E.} \sur{Wortley}}

\affil[1]{\orgdiv{School of Physics and Astronomy}, \orgname{University of Birmingham}, \orgaddress{\street{Edgbaston}, \city{Birmingham}, \postcode{B15 2TT}, \country{UK}}}

\affil[2]{\orgdiv{Institute for Gravitational Wave Astronomy}, \orgname{University of Birmingham}, \orgaddress{\street{Edgbaston}, \city{Birmingham}, \postcode{B15 2TT}, \country{UK}}}

\affil[3]{\orgdiv{School of Physics and Astronomy}, \orgname{University of Leicester}, \orgaddress{\street{University Road}, \city{Leicester}, \postcode{LE1 7RH}, \country{UK}}}

\affil[4]{\orgname{Eureka Scientific}, \orgaddress{\city{Oakland}, \state{CA}, \country{USA}}}

\affil[5]{\orgdiv{Aix Marseille Univ.}, \orgname{CNRS, CNES, LAM}, \orgaddress{\city{Marseille}, \postcode{F-13388}, \country{France}}}

\affil[6]{\orgname{Cosmic Dawn Center (DAWN)}, \orgaddress{\country{Denmark}}}

\affil[7]{\orgdiv{Niels Bohr Institute}, \orgname{University of Copenhagen}, \orgaddress{\street{Jagtvej 155A}, \city{Copenhagen}, \postcode{DK-2200}, \country{Denmark}}}

\affil[8]{\orgdiv{Centro Astronómico Hispano en Andalucía}, \orgname{Observatorio de Calar Alto}, \orgaddress{\street{Sierra de los Filabres}, \city{Gérgal}, \postcode{E-04550}, \state{Almería}, \country{Spain}}}

\affil[9]{\orgname{Institut de Radio Astronomie Millimétrique (IRAM)}, \orgaddress{\street{300 rue de la Piscine}, \postcode{38406}, \state{Saint Martin d’Héres}, \country{France}}}

\affil[10]{\orgdiv{Department of Astrophysics/IMAPP}, \orgname{Radboud University}, \orgaddress{\street{PO Box 9010}, \postcode{6500 GL Nijmegen}, \country{The Netherlands}}}

\affil[11]{\orgdiv{Department of Physics}, \orgname{University of Warwick}, \orgaddress{\street{}, \city{Coventry}, \postcode{CV4 7AL}, \country{UK}}}

\affil[12]{\orgdiv{David A. Dunlap Department of Astronomy and Astrophysics}, \orgname{University of Toronto}, \orgaddress{\city{Toronto}, \postcode{M5S 3H4}, \state{ON}, \country{Canada}}}

\affil[13]{\orgname{E. Kharadze Georgian National Astrophysical Observatory}, \orgaddress{\street{Mt. Kanobili}, \city{Abastumani}, \postcode{0301}, \state{Adigeni}, \country{Georgia}}}

\affil[14]{\orgdiv{Department of Physics \& Astronomy}, \orgname{Louisiana State University}, \orgaddress{\city{Baton Rouge}, \postcode{70803}, \state{LA}, \country{USA}}}

\affil[15]{\orgdiv{Space Science Data Center (SSDC)}, \orgname{Agenzia Spaziale Italiana (ASI)}, \orgaddress{\city{Roma}, \postcode{00133}, \country{Italy}}}

\affil[16]{\orgdiv{MIFT Department}, \orgname{University of Messina}, \orgaddress{\street{Via F.S. D'Alcontres 31}, \city{Messina}, \postcode{98166}, \country{Italy}}}

\affil[17]{\orgdiv{Department of Physics and Astronomy}, \orgname{Clemson University}, \orgaddress{\city{Clemson}, \postcode{29634-0978}, \state{SC}, \country{USA}}}

\affil[18]{\orgdiv{Centre for Astrophysics and Cosmology, Science Institute}, \orgname{University of Iceland}, \orgaddress{\street{Dunhagi 5}, \city{Reykjavík}, \postcode{107}, \country{Iceland}}}

\affil[19]{\orgname{Th\"uringer Landessternwarte Tautenburg}, \orgaddress{\street{Sternwarte 5}, \city{Tautenburg}, \postcode{07778}, \country{Germany}}}

\affil[20]{\orgdiv{School of Physics and Centre for Space Research}, \orgname{University College Dublin}, \orgaddress{\street{Belfield}, \city{Dublin}, \postcode{4}, \country{Ireland}}}

\affil[21]{\orgdiv{Astrophysics Research Centre, School of Mathematics and Physics}, \orgname{Queen’s University Belfast}, \orgaddress{\city{Belfast}, \country{UK}}}

\affil[22]{\orgdiv{Anton Pannekoek Institute of Astronomy}, \orgname{University of Amsterdam}, \orgaddress{\street{Science Park 904}, \city{Amsterdam}, \postcode{1098 XH}, \country{The Netherlands}}}

\affil[23]{\orgname{Max-Planck-Institut f\"ur Extraterrestrische Physik}, \orgaddress{\street{Giessenbachstra\ss{}e 1}, \city{Garching}, \postcode{85748}, \country{Germany}}}

\affil[24]{\orgname{INAF – Osservatorio di Astrofisica e Scienza dello Spazio}, \orgaddress{\street{Via Piero Gobetti 101}, \city{Bologna}, \postcode{I-40129}, \country{Italy}}}

\affil[25]{\orgdiv{The Oskar Klein Centre, Department of Astronomy}, \orgname{Stockholm University}, \orgaddress{\city{Stockholm}, \postcode{SE-106 91}, \country{Sweden}}}

\abstract{While the progenitors of many long gamma-ray bursts (GRBs) are well established as the core collapse of very massive, envelope-stripped, rapidly-rotating stars, it has been suggested that bursts at the extremely long end of the duration distribution may be a separate population of `ultra-long' GRBs. With durations of thousands of seconds or more, these bursts are difficult to reconcile with the engine timescales expected from the compact Wolf-Rayet stars that are typically assumed to produce more `standard' long GRBs. Here, we present observations of GRB~220706A, where X-ray follow-up reveals flaring episodes that last until $\approx 51$ days after trigger. At the measured redshift of $z = 0.8577$, this corresponds to 27 days in the GRB rest frame, and represents the latest central engine activity ever observed in a GRB by a margin of $\approx 21$ rest-frame days. We also identify a likely supernova (SN) which, when accounting for the inferred optical extinction of $0.9 \leq A_V \leq 3.6$ mag (constrained by indirect arguments), has a peak absolute magnitude of $M_r \leq -20.25$, resembling the energetic SN\,2011kl found accompanying ultra-long GRB\,111209A, and consistent with super-luminous SNe. We discuss the implications GRB~220706A has for ultra-long GRB progenitor models and possible powering mechanisms for the extremely late central engine activity.}




\maketitle


GRB~220706A was first detected by the Burst Alert Telescope (BAT) on board the Neil Gehrels \textit{Swift} Observatory (hereafter \textit{Swift}) at 16:09:14 UT on 2022 July 6. \textit{Swift}'s X-ray Telescope (XRT) and Ultra-Violet Optical Telescope (UVOT) began pointed observations $102.8$\,s after trigger but did not detect an optical counterpart. We obtained further X-ray observations from the Neutron star Interior Composition Explorer (NICER) between 2022 July 7 and 2022 August 16, and from the \emph{Chandra} X-ray Observatory on 2022 August 16. In addition to the automatic triggered response, we obtained late-time \emph{Swift}/XRT observations between 2022 August 7 and 2022 October 3. X-ray observations are shown in Table~\ref{tab:opt_phot} except \emph{Swift}, which can be found on the UK \emph{Swift} Science Data Centre (UKSSDC; \cite{Evans07,Evans09})\footnote{\url{https://www.swift.ac.uk/xrt\_curves/01114937/}}.

Optical and near-infrared follow-up of GRB\,220706A began as soon as the field was visible from La Palma, Spain (about 0.52 days after the GRB), initially with the Nordic Optical Telescope (NOT) and later with the Gamma-Ray burst Optical and Near-infrared Detector (GROND) at La Silla, Chile, the ESO Very Large Telescope (VLT) FORS2 instrument at Cerro Paranal, Chile, and the 10.4-meter Gran Telescopio Canarias (GTC) in La Palma. Imaging observations continued until $80$ days after the GRB trigger.

\begin{figure}
  \centering\includegraphics[width=\columnwidth]{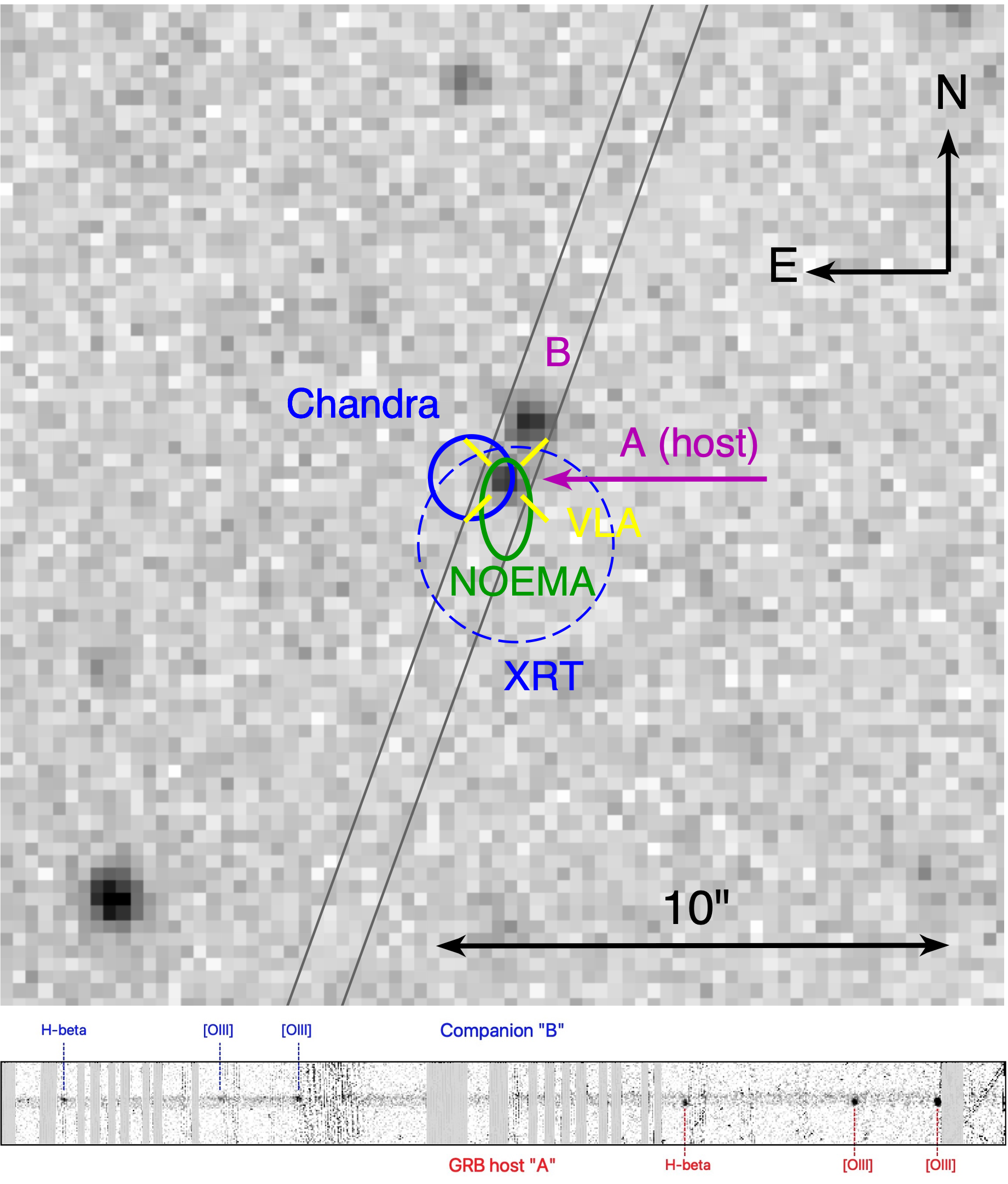}
  \caption{\textbf{Top:} The field of GRB\,220706A as captured in a VLT/FORS2 $R$-band image (approximately $20''$ in side). The orientation of the GTC/OSIRIS slit is also indicated. \textbf{Bottom:} The optical spectrum of the field of GRB\,220706A, including the host galaxy (A) at a redshift of $z=0.8577\pm0.0005$ and the nearby galaxy (B) at $z=0.7207\pm0.0005$. Wavelengths with strong atmospheric lines, resulting in prominent residuals, have been masked in grey to improve visibility.}\label{fig:opticalfield}
\end{figure}

Imaging revealed two marginally extended objects ($\approx 1''$ PSF under a $0.65''$ seeing), marked ``A'' and ``B'' in Figure~\ref{fig:opticalfield}. Object A has a magnitude of $r = 25.52 \pm 0.06$ and lies at coordinates $\mbox{RA(J2000)} = \mbox{00:02:25.056}$, $\mbox{Dec(J2000)} = -\mbox{16:24:38.53}$, within the XRT error circle. Object B lies just to the north of the XRT error circle, though is plausibly consistent with it. Although no optical transient was initially identified, image subtraction of the early observations was performed using deep, late templates (see Section~\ref{subsec:VLT}) and the optical counterpart to GRB\,220706A was discovered at $\mbox{RA(J2000)} = \mbox{00:02:25.08}(\pm 0.01''$), $\mbox{Dec(J2000)} = -\mbox{16:24:38.5}(\pm 0.20''$). The resulting host-subtracted photometry is listed in Table \ref{tab:opt_phot}.

Spectroscopy of the host galaxy (A) and its companion (B) was obtained using GTC/OSIRIS at a mean epoch of 2022 August 29 at 03:07 UT (53.46 days after the burst). The orientation of the slit is indicated in Figure~\ref{fig:opticalfield}. The combined 2D spectrum, shown in the lower panel of Figure~\ref{fig:opticalfield}, reveals a faint trace corresponding to the two objects. Both appear as star-forming galaxies with prominent emission lines due to H$\beta$ and [OIII]. Using these line identifications, we measure redshifts of $z=0.8577\pm0.0005$ for the GRB host (galaxy A), and of $z=0.7207\pm0.0005$ for the nearby galaxy (B) -- thus showing that the two objects are not physically related.

Longer wavelength follow-up of GRB\,220706A was obtained with the NOEMA interferometer on 2022 July 8 and with the Karl G. Jansky Very Large Array (VLA) on 2023 October 5. The positions obtained from these observations corroborate the association of GRB\,220706A with galaxy A (Figure~\ref{fig:opticalfield}). Measured flux densities are given in Table~\ref{tab:opt_phot}.

\begin{table}
    \centering
    \begin{tabular}{lcccc}
    \hline\hline
         $\Delta T$ &   Telescope/camera & Filter & Exp & Magnitude \\
         (day) & & & (s) & (AB) \\
    \hline
        0.52  & NOT/ALFOSC & $R$ &  900 &$>24.00$\\
        0.73 & 2.2m/GROND & $g'$ & 1600 & $>25.19$ \\
        0.73 & 2.2m/GROND & $r'$ & 1600 & $>23.39$ \\
        0.73 & 2.2m/GROND & $i'$ & 1600 & $>24.10$ \\
        0.73 & 2.2m/GROND & $z'$ & 1600 & $>23.77$ \\
        0.73 & 2.2m/GROND & $J$ & 1600 & $>21.65$ \\
        0.73 & 2.2m/GROND & $H$ & 1600 & $>20.92$ \\
        0.73 & 2.2m/GROND & $K$ & 1600 & $>19.88$ \\ 
        1.51  & GTC/EMIR   & $J$ & 2450 &$> 23.26$ \\
        7.53  & NOT/ALFOSC & $R$ & 1800 &$>23.55$ \\
        18.5 & NOT/ALFOSC & $R$ & 1800 & $24.39\pm0.29$ \\
        33.7 & VLT/FORS2  & $R$ &  900 & $24.51\pm0.10$\\
        34.5 & GTC/OSIRIS & $i$ & 1260 & $24.13\pm0.30$\\
        34.5 & GTC/OSIRIS & $r$ & 1200 & $24.97\pm 0.13$\\
        34.5 & GTC/OSIRIS & $z$ & 1200 & $23.72\pm0.10$\\ 
        45.6 & VLT/FORS2  & $R$ &  900 & $24.99\pm0.11$  \\
        51.4 & NOT/ALFOSC & $R$ & 3000 & $>24.40$ \\
        53.4 & GTC/OSIRIS & $i$ &  900 & $> 24.11$\\
        53.4 & GTC/OSIRIS & $z$ &  900 & $>24.89$\\
        79.6 & VLT/FORS2  & $R$ &  900 & $>25.67$ \\
        \hline\hline
        $\Delta T$ & Telescope & Band & Exp & Flux \\
         (day) & & & (ks) & (erg\,cm$^{-2}$\,s$^{-1}$) \\
        \hline
        0.75 & NICER & 0.3-10\,keV & 61.9 & 2.24$^{+0.52}_{-0.42} \times 10^{-12}$ \\
        1.18 & NICER & 0.3-10\,keV & 11.5 & 1.70$^{+0.12}_{-0.11} \times 10^{-11}$ \\
        1.36 & NICER & 0.3-10\,keV & 19.2 & 5.89$^{+0.87}_{-1.52} \times 10^{-12}$ \\
        1.50 & NICER & 0.3-10\,keV & 5.6 & 9.77$^{+0.72}_{-0.36} \times 10^{-13}$ \\
        40.9 & \textit{Chandra} & 0.3-10\,keV & 15.0 & $3.09^{+0.93}_{-1.16} \times10^{-14}$ \\
        \hline\hline
        $\Delta T$ & Observatory & Frequency & Exp & Flux density \\
         (day) & & (GHz) & (hr) & ($\mu$Jy) \\
        \hline
        1.51 & NOEMA & 74.3 & 0.82 & $<156$ \\
        1.51 & NOEMA & 89.7 & 0.82 & $97 \pm 26$ \\
        1.62 & NOEMA & 136.3 & 1.40 & $<201$ \\
        1.62 & NOEMA & 151.7 & 1.40 & $<252$ \\
        455.7 & VLA & 3 & 0.78 & $53.2 \pm 12.8$ \\
    \hline\hline
    \end{tabular}
    \caption{Observations of the GRB\,220706A afterglow, showing host-subtracted optical photometry, X-ray (excl. \emph{Swift}), and long-wavelength detections. $\Delta T$ is the time since the \textit{Swift}/BAT burst trigger in the observer frame. AB magnitudes for the transient emission were derived using a host galaxy reference template taken $700$ (FORS2) and $1096$ (GTC) days post-burst to ensure accurate host subtraction. Upper limits are given at the $3\sigma$ level.}
    \label{tab:opt_phot}
\end{table}

\section{Results}\label{sec:results}

\subsection{GRB 220706A as an ultra-long GRB}\label{sec:ULGRB}

The measured duration of GRB\,220706A in the \emph{Swift}/BAT bandpass is $t_{90} = 87 \pm 18$\,s \cite{Stamatikos22}, consistent with the peak of the $t_{90}$ distribution presented in \cite{Lien16}. It is therefore not obviously distinct from the population of regular long GRBs by this metric. However, the $t_{90}$ parameter measures the middle 90\% of the gamma-ray emission duration in the detector bandpass rather than the full duration of the central engine. Since most of the variable emission episodes in GRB\,220706A were observed in X-rays (see Section~\ref{sec:flares}), we instead utilise the $t_{\rm burst}$ metric \cite{Zhang14}, which defines the duration of the central engine as the last point at which the light curve is in the steep decay phase (with a measured temporal index of $t^{-3}$ or steeper) or the gamma-ray $t_{90}$; whichever is greater.

We measure $t_{\rm burst}$ for GRB\,220706A and 549 other GRBs from the UKSSDC that pass data quality cuts (see Section~\ref{sec:tburst}). Within this sample, GRB\,220706A has the 8$^{\rm th}$ longest $t_{\rm burst}$ duration with a value of $t_{\rm burst} = 10^{4.75}$\,s. This makes it a clear member of the ultra-long population, with an engine duration in excess of those seen in the archetypal ultra-long GRBs\,111209A ($t_{\rm burst} = 10^{4.68}$~s), 130925A ($t_{\rm burst}= 10^{4.45}$~s) and 121027A ($t_{\rm burst} = 10^{4.36}$~s).

\subsection{High-energy evolution}\label{sec:flares}

\begin{figure}
    \centering
    \includegraphics[width=\columnwidth]{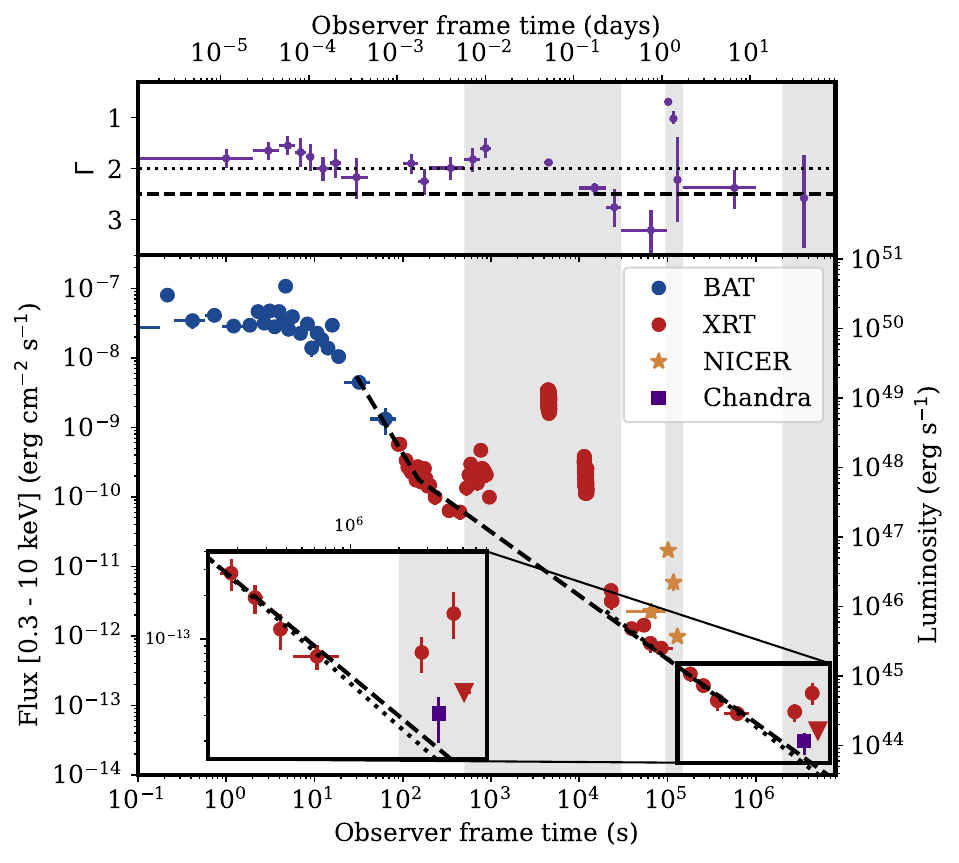}
    \caption{The X-ray and $\gamma$-ray light curve of GRB\,220706A. We use the S/N = 5 non-spectrally-evolving BAT light curve extrapolated to the XRT band, which was taken directly from the UKSSDC Burst Analyser \cite{Evans10}. The dashed line shows a broken power-law model fit to the data between 30\,s and $10^6$\,s. The dotted line shows a power-law model fit to the data between $3\times 10^4$\,s and $10^6$\,s. Their close agreement suggests there is a small section of afterglow-dominated emission before the start of the main flare at $\sim 500$\,s. In both cases, the flaring epochs marked in grey are excluded from the fit. The top panel shows the time-evolving photon index from simple power-law fits to the data. Lower photon indices (harder spectra) are towards the top.}
    \label{fig:xrays}
\end{figure}

The combined X-ray and $\gamma$-ray light curve of GRB\,220706A is shown in Figure~\ref{fig:xrays}. By excluding epochs with flaring, the data from 30\,s after trigger can be fit with a broken power-law model, suggesting that there may be a brief period of afterglow-dominated emission before the onset of the main flaring episode at $\sim 500$\,s. The best fit ($\chi^2$/dof = 0.87) has temporal indices of $\alpha_1 = 2.12 \pm 0.14$ and $\alpha_2 = 0.92 \pm 0.01$ around a break time of $t_b = 150 \pm 13$\,s and a normalisation (the flux at 1s) of $N = (7.2 \pm 4.5) \times 10^{-6}$\,erg\,cm$^{-2}$\,s$^{-1}$. We also fit a power-law model from $3 \times 10^4$\,s, finding $\alpha = 0.98 \pm 0.07$ and $N = (3.6 \pm 2.9) \times 10^{-8}$\,erg\,cm$^{-2}$\,s$^{-1}$ ($\chi^2$/dof = 1.02), in good agreement with the broken power-law fit.

X-ray observations continue to show episodes of flaring until very late times ($t > 10^6$\,s). The \textit{Chandra} epoch is marginally consistent with the extrapolated power-law, lying $1.48\sigma$ above it, while the XRT observations taken before and after are at higher flux, lying $2.80\sigma$ and $2.84\sigma$ above the nominal afterglow. Corrected for redshift, this last epoch of flaring suggests that the central engine of GRB\,220706A is still active at $27.25^{+1.95}_{-0.03}$ rest-frame days after trigger. This is $\approx20.6$ rest-frame days later than the flares seen in GRB\,210204A, which were previously claimed to be the most delayed ever observed \cite{Kumar22}.

We also fit a time-resolved spectral series to the data (Section~\ref{sec:spectra}) which is shown in the top panel of Figure~\ref{fig:xrays} and Extended Data Table~\ref{tab:spec_series}. The variability of the measured photon indices demonstrates the frequent engine activity of GRB\,220706A. Very little of the light curve appears to be afterglow dominated. The main region of smooth power-law decay, indicating afterglow-only emission, lies between $\sim 10^5$ and $10^6$s, where the photon index of $\Gamma = 2.38^{+0.41}_{-0.36}$ is poorly constrained due to low flux. Under standard GRB afterglow assumptions \cite{Sari98}, the measured photon indices and decay rate of the X-ray afterglow imply GRB\,220706A exploded in a constant density (ISM-like) environment and the synchrotron cooling break is below the X-ray bandpass (see Section~\ref{sec:inferences}).

The brightest flare detected by XRT occurred between 4000 and 5000\,s after trigger, with a $0.3$ -- $10$\,keV flux of $\mbox{a few} \times 10^{-9}$\,erg\,cm$^{-2}$\,s$^{-1}$. The high count rate allows us to further sub-divide our spectra to investigate the engine driving the flare (see Section~\ref{sec:XRTflare}). The best-fitting model is a power-law with a high-energy exponential cutoff of a few keV above a photon index fluctuating around $\Gamma \sim 1$, in contrast to the value of $\Gamma \sim 2$ found when using a power-law fit (e.g. Figure~\ref{fig:xrays}). This finding implies the presence of a low-energy spectral break (e.g. \cite{Ravasio19}) or a thermal emission component in the bandpass, although the narrow energy range of XRT precludes a definitive identification.

More than a day after the GRB trigger, NICER detected a flare with a flux of a few $\times 10^{-11}$\,erg\,cm$^{-2}$\,s$^{-1}$. This flare is unusually late; most X-ray flares occur within 1000s of the GRB trigger \cite{Swenson14}. A time-resolved spectral analysis of the NICER data around this flare (see Section~\ref{sec:NICERflare}) shows that an absorbed power-law model provides an adequate description of the spectra in all epochs except Epoch~2, corresponding to the peak of the flare. In this interval, we identify a systematic Gaussian-like residual centered at $\sim$2.5~keV in the observer frame ($5.3^{+0.3}_{-0.3}$~keV in the rest frame). A full discussion of the significance of this feature is presented in Section~\ref{sec:NICERflare}.

\subsection{Broad-band afterglow and optical extinction constraints}\label{sec:SED}

\begin{figure}
    \centering
    \includegraphics[width=\columnwidth]{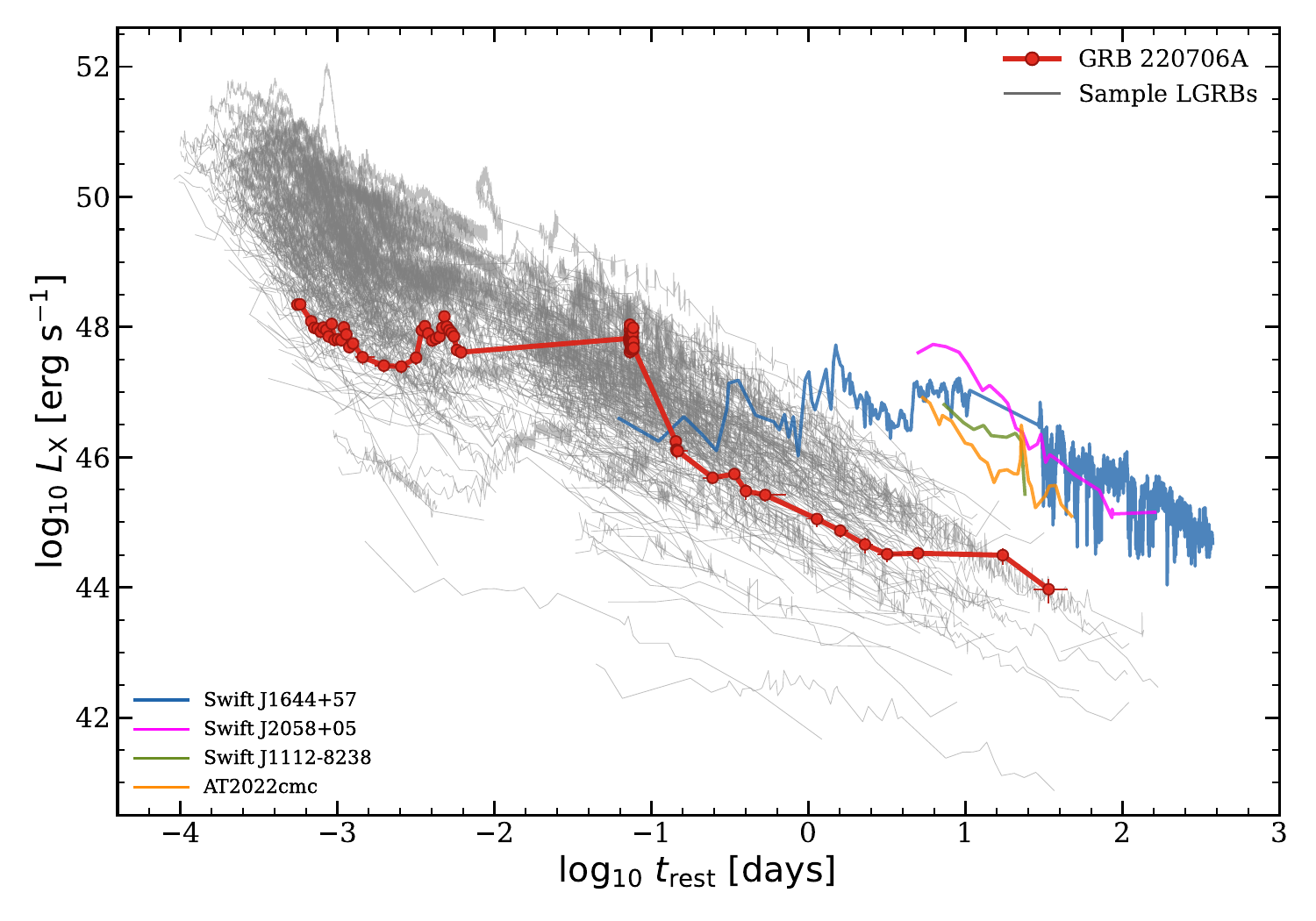}
    \caption{The X-ray afterglow of GRB\,220706A (red) shown in the context of observed X-ray afterglow light curves in the \textit{Swift} XRT energy band (0.3 -- 10 keV) for a sample of 323 LGRBs ($t_{90}>2$~s) with known redshift, together with the X-ray light curves of the relativistic TDEs Swift J1644+57, Swift J2058+05, Swift J1112$-$8238, and AT2022cmc }
    \label{fig:Lum_xray}
\end{figure}

To place the X-ray afterglow of GRB\,220706A in context, we compare its light curve to a comprehensive sample of LGRBs with known redshifts (Figure~\ref{fig:Lum_xray}). GRB\,220706A possesses a relatively low luminosity X-ray afterglow, consistently falling below the median luminosity of the broader LGRB sample until late times, when its flaring behavior makes it one of the brightest events.

The deepest early $r$-band observation, obtained by NOT/ALFOSC at $0.52$ days post-burst, yields an upper limit of $r > 24.00$ mags, equivalent to a flux density of $F_r < 0.91$\,$\mu$Jy. This is remarkably faint; under standard afterglow assumptions, the spectral slope between X-rays and optical is expected to be $\beta_{OX} \geq 0.5$ \cite{Sari98}. The $r$-band upper limit implies that the X-ray afterglow should have a 1\,keV flux density of $F_X < 4.08 \times 10^{-2}$\,$\mu$Jy at this epoch, but the interpolated model fit to the X-ray afterglow (see Section~\ref{sec:flares}) indicates a 1\,keV flux density of $F_X \approx 0.14$\,$\mu$Jy. Comparing the interpolated X-ray flux to the optical upper limit, we find that $\beta_{OX} < 0.30$. This value satisfies the criterion for classifying GRB\,220706A as a dark burst \cite{2004ApJ...617L..21J}.
 
Several factors could explain the dark nature of the burst. One likely explanation is substantial dust extinction within the host galaxy, which potentially obscures the afterglow emission (e.g. \cite{2011A&A...534A.108K}). Alternatively, the uniformly faint afterglow across all wavelengths may indicate a low-density environment surrounding the burst \cite{2009AJ....138.1690P}. In support of the former interpretation, we infer a significant amount of host galaxy absorption from the late XRT spectrum fit (Section~\ref{sec:flares}), with a hydrogen column density of $N_{H,X} =(2.43 \pm 0.29) \times 10^{22}$\,cm$^{-2}$. Although $A_{V}/N_{H,X}$ varies by nearly a factor of ten across different \textit{Swift} bursts, using the empirical relation from \cite{2007MNRAS.377..273S} we estimate $A_{V} \approx 3.6$ mag from our measured value of $N_{H,X}$. This high line of sight extinction could explain the dark optical nature of the GRB\,220706A afterglow.

\begin{figure}
    \centering
    \includegraphics[width=\columnwidth]{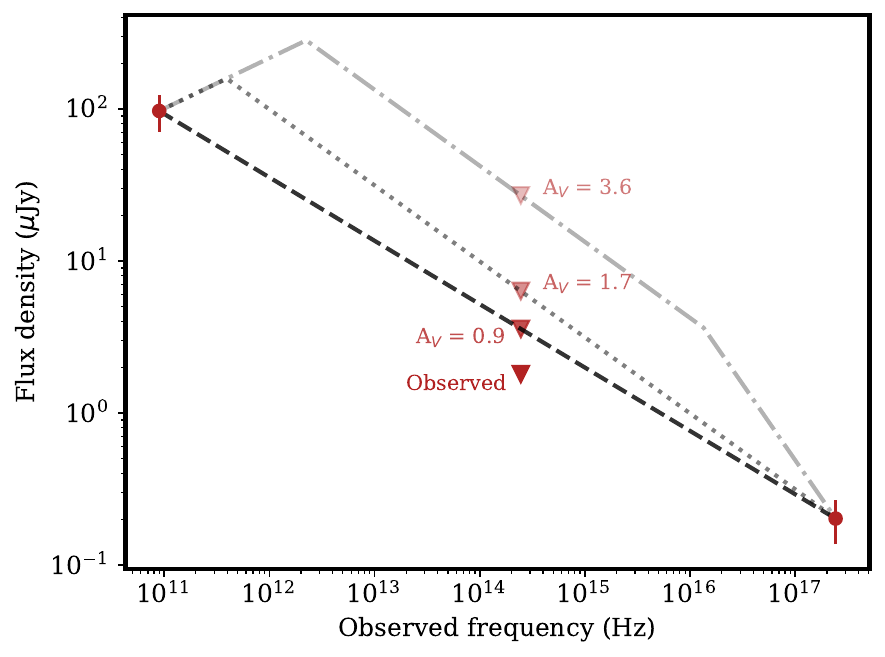}
    \caption{The broad-band SED of the GRB\,220706A afterglow at 1.51 observer frame days. The data shown are the detection from NOEMA, the J-band upper limit from GTC, and the average flux of the XRT datapoints either side of the target epoch. The X-ray datapoint has been corrected for absorption by a factor of $f_{\rm unabs}/f_{\rm obs} = 3$, based on the time-averaged spectral fit from the UKSSDC. The solid triangle shows the observed GTC upper limit, while increasingly faint triangles represent upper limits after correcting for host extinction of: $A_V = 0.9$, to match the measured spectral slope ($\beta = 0.42 \pm 0.04$, dashed line); $A_V = 1.7$, to match the shallowest expected theoretical spectral slope ($\beta = 0.5$, dotted line); and $A_V = 3.6$, estimated from $A_V/N_{H,X}$ scaling relations for GRB hosts \cite{2007MNRAS.377..273S}, with a representative model that assumes the synchrotron cooling break lies between the infra-red and X-ray bands, as found in Section~\ref{sec:flares}. $R_V = 3.1$ is assumed in all cases.}
    \label{fig:SED}
\end{figure}

We can place a robust lower limit on $A_V$ by constructing a spectral energy distribution (SED) from the XRT, GTC and NOEMA data taken at a consistent epoch of $\sim 1.51$ observer-frame days after trigger (Figure \ref{fig:SED}). Since GRB afterglows are powered by synchrotron radiation, the broad-band spectrum will be a power-law or broken power-law \cite{Sari98}. A simple power-law model between the sub-mm and X-ray data therefore defines a minimum $A_V$ at optical/IR frequencies; the extinction correction must make the GTC observations consistent with this line, and any break in the spectrum introduces a steeper spectral component that can only \textit{increase} the required $A_V$. We find $\beta = 0.42 \pm 0.04$ between the X-rays and sub-mm observations, which requires $A_V > 0.9$ for an assumed $R_V = 3.1$. Furthermore, in the case where $\beta = 0.5$, the theoretically expected minimum value, a peak in the spectrum is required above the NOEMA bandpass (Figure~\ref{fig:SED}). Under this assumption, the required extinction would be $A_V > 1.7$.

Finally, it is possible that the spectrum is intrinsically shallow. The electron energy power-law index is typically assumed to be $2 \leq p \leq 3$, with the lower bound corresponding to $\beta = 0.5$. However, values of $p < 2$ have been inferred in previous studies (e.g. \cite{Gompertz18,Anderson25}), most notably in GRB\,221009A, the brightest GRB ever observed \cite{Levan23b}. On the other hand, as discussed in Section~\ref{sec:flares}, the X-ray data of the GRB\,220706A afterglow are broadly consistent with $p \approx 2$.

\subsection{Identification of a likely supernova component}\label{sec:SN}

Optical emission from GRB\,220706A was first detected at $\sim$17 days post-burst at a flux an order of magnitude in excess of expectations for the afterglow based on early non-detections and its evolution X-rays. This is consistent with the emergence of a SN component. To explore this possibility, we fit the $r$-band light curve with a combined model comprised of a decaying power law ($F_{\nu} \propto t^{-\alpha}$) and a SN template based on SN\,1998bw \cite{Galama98,2011AJ....141..163C}.

\begin{figure}
    \centering
    \includegraphics[width=\columnwidth]{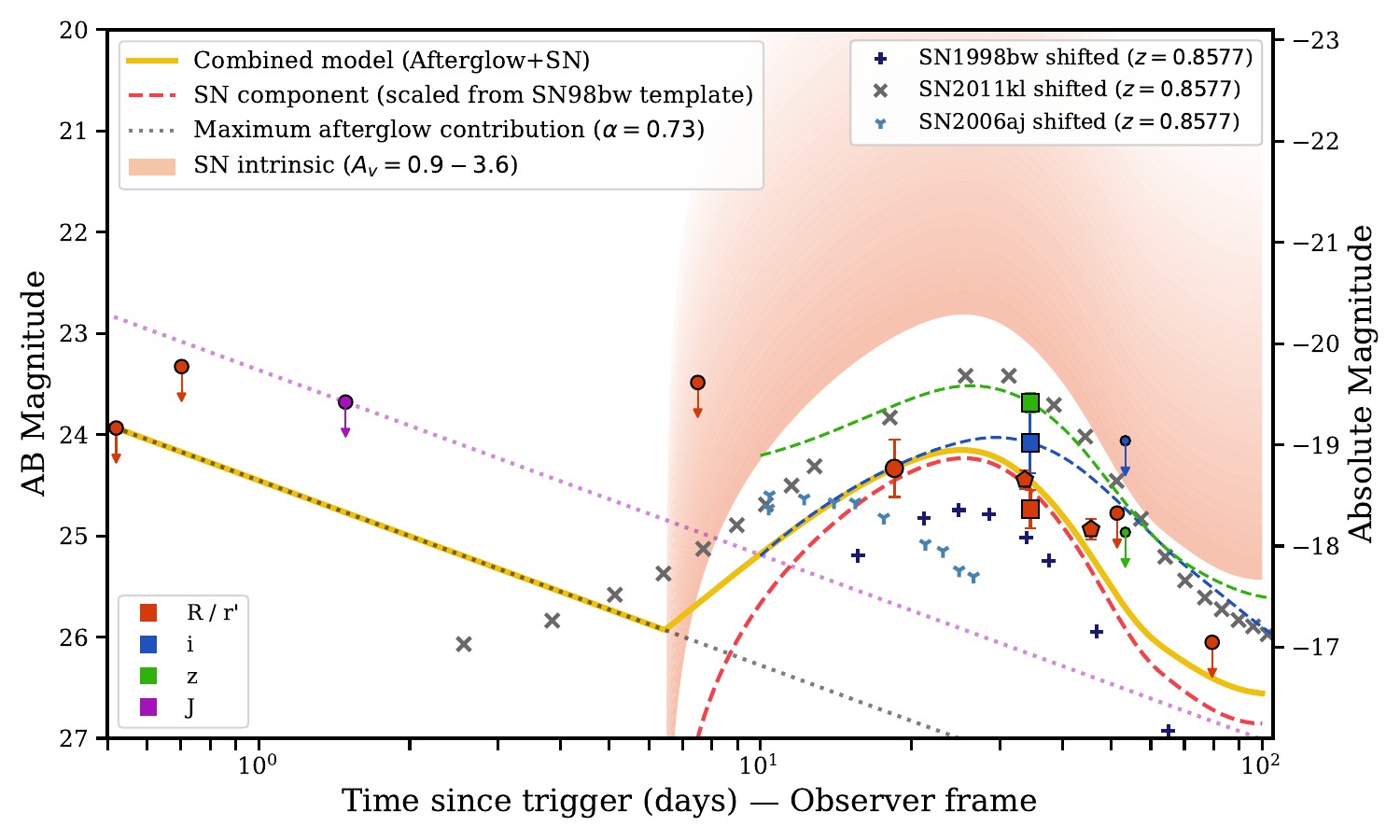}
    \caption{Host-subtracted optical-NIR photometry of GRB\,220706A in the observer frame. The solid yellow line represents the combined afterglow+SN model and is fitted to the $R/r'$-band data. The dotted lines show the maximum allowed afterglow contribution, constrained by the $R/r'$ and $J$-band upper limits. The red dashed curve shows the best-fitting SN component in the $R/r'$ band, modeled using the SN\,1998bw $U$-band template. The best fit of $k=1.58 \pm 0.23$ corresponds to a peak that is $\sim0.5$ mag brighter than SN\,1998bw. The dashed blue and dashed green curves show SN-only fits to the $i$- and $z$-band data, respectively, obtained by scaling the SN 1998bw $V$- and $B$-band templates with the stretch factor fixed to 1. The shaded region represents the range of possible SN $R/r'$-band magnitudes after correcting for host extinction of $A_V =0.9 - 3.6$, derived from broadband spectral constraints (see Section~\ref{sec:SED}). Template light curves of SN\,1998bw ($+$), SN\, 2006aj ($\Upsilon$) and SN\,2011kl ($\times$), converted to $z = 0.8577$, are shown for comparison.}
    \label{fig:opt_LC_SN}
\end{figure}

The afterglow component was pinned to the early NOT upper limit and given a temporal index of $\alpha = 0.73$ based on the constraints obtained from the X-ray afterglow (Section~\ref{sec:flares}), which indicated a constant density ISM and the synchrotron cooling frequency lying between the optical and X-ray bands. The SN\,1998bw U-band template (which corresponds to the $r$-band at $z = 0.8577$) was interpolated to the observed $r$-band epochs of GRB\,220706A using a cubic spline and fit with a luminosity scaling factor (k) with a stretch factor (s) fixed to 1. The best fit gave $k=1.58\pm0.23$, corresponding to a SN that is $\sim 0.5$ magnitudes brighter at peak than SN\,1998bw. This is equivalent to a peak absolute magnitude of $M_r = -18.85$. The best-fit model is shown in Figure~\ref{fig:opt_LC_SN} alongside its $i$- and $z$-band counterparts and three GRB-associated SN templates: SN\,1998bw \cite{2011AJ....141..163C}, SN\,2006aj \cite{2006A&A...454..503S} and SN\,2011kl \cite{Greiner15}.

The observer $r$-band probes an effective rest-frame wavelength of 3540\AA{} at $z = 0.8577$, where the host galaxy extinction would be in the range of 1.4 -- 5.7 mags, corresponding to $0.9 \leq A_V \leq 3.6$ (Section~\ref{sec:SED}). The peak absolute magnitude of the SN is therefore $M_r \leq -20.25$ \emph{even in the rest-frame UV}, which is approaching magnitudes associated with the super-luminous SN category \cite{Gal-Yam19,Nicholl21,Gomez24}, and would exceed the nominal threshold of $M_r \leq -21$ in the case of $A_V = 1.7$ ($M_r = -21.56$), which corresponds to the shallowest expected spectral slope from synchrotron theory (Figure~\ref{fig:SED}). This would be the first detection of a GRB and a SLSN from a single event (see e.g. \cite{Coppejans18,Margutti18,Eftekhari21} for previous constraints).

However, an important caveat to the SN interpretation is that the implied optical rise begins at around 5 rest-frame days, during a gap in the X-ray coverage. The SN peak is broadly consistent with the late X-ray flaring epoch. It is therefore possible that the inferred SN is the optical counterpart to the X-ray flaring, which happens to mimic the evolution and luminosity of a bright SN similar to rare events seen alongside other GRBs \cite{Greiner15,Kann24}. Optical flares with SN luminosities have previously been observed in luminous fast blue optical transients (FBOTs), although with durations of minutes rather than days to weeks \cite{Ho23}. These were attributed to non-thermal processes, likely associated with a relativistic jet.

\section{Discussion}\label{sec:discussion}

\subsection{The origin of the extremely late emission}\label{sec:engines}

The continued flaring in GRB\,220706A at 27 rest-frame days indicates that an emission process in addition to the GRB afterglow contributes to the light curve. This is unprecedented in the GRB archives, though this is in part due to a lack of coverage at these epochs.

\subsubsection*{Accretion}\label{sec:accretion}

In the canonical GRB engine, the burst is powered by the infall of the stellar envelope, which is accreted onto a remnant black hole formed by the collapse of the core. The timescale for accretion and the duration of the GRB is therefore a function of the rotation and density profile of the progenitor star \cite{Kumar08}. Under the assumption of free-fall collapse, the accretion timescale can be approximated by the free-fall time of the stellar material,
\begin{equation}
    t_{\rm ff} = \frac{\pi r^{3/2}}{(2GM)^{1/2}},
\end{equation}
where $r$ and $M$ are the stellar radius and mass, respectively. For accretion onto the black hole to still occur at 27 days after collapse, a $\sim 10$\,$M_{\odot}$ star with a $\sim 10^{13}$\,cm radius is required. This implies a supergiant progenitor star, as has previously been suggested for ultra-long GRBs (e.g. \cite{Quataert12,Levan14}). Models for the free-fall accretion rate of supergiant stars \cite{Woosley02} show an uptick in available accretion power at timescales and luminosities appropriate for the flares seen in GRB\,220706A (see figure 1 of \cite{Quataert12}). However, the presence of a SN invalidates the assumption of continuous free-fall at late times.

An alternative is the massive but diffuse stars predicted to occur as a result of pulsational pair instability. These have previously been suggested as candidates for ultra-long GRB progenitor stars due to their extremely extended envelopes (up to $\sim 10^{14}$\,cm \cite{Marchant20}), with the associated luminous SN produced by the high temperatures preserved by the large progenitor radius \cite{Moriya20}. However, while they have been shown to provide a possible explanation for GRB\,111209A / SN\,2011kl, the model has not been applied to flaring episodes as late as those seen in GRB\,220706A.

\subsubsection*{Magnetar Activity}\label{sec:magnetar}

Under the magnetar model, the rotational energy of a post-collapse neutron star with an extremely strong dipole magnetic field is deposited into the expanding ejecta at a rate determined by the field strength (B) and neutron star spin period (P) \cite{Zhang01}. This provides a very natural long-lived engine, and the model is commonly used to explain X-ray plateaus in the afterglows of both long and short GRBs (e.g. \cite{Rowlinson13,Gompertz13}). A magnetar is an appealing solution for GRB\,220706A due to the success of the model in matching ultra-long GRB\,111209A / SN\,2011kl \cite{Greiner15,Metzger15,Gompertz17}.

The high-energy observations of GRB\,220706A suggest that magnetar solutions can be found for values of $P \leq 2$\,ms and $5\times 10^{14} \leq B \leq 10^{15}$\,$G$, consistent with the expected ultra-long GRB parameter space outlined in \cite{Metzger15} and values derived for ultra-long GRB 111209A \cite{Gompertz17}. However, such solutions appear either fine tuned in order to hide the characteristic dipole plateau beneath the flaring episodes, or qualitatively inconsistent with the expected energy partition between the GRB and SN when compared to GRB 111209A / SN\,2011kl \cite{Metzger15}.

Flares of similar luminosity have been observed alongside other magnetar-associated transients like SLSNe \cite{Levan13} and Fast Blue Optical Transients (FBOTs; \cite{Margutti19}), but were attributed to ionization breakout, requiring a separate emission site to the GRB afterglow. Magnetars can also drive giant flares with luminosities comparable to those seen in GRB\,220706A through global crust failures and magnetic reconnection (e.g. \cite{Hurley05}). However, the typical timescales of such flares are of the order of seconds to minutes, and nascent magnetars are not expected to have fully formed crusts at timescales relevant to GRB\,220706A.

\subsubsection*{CSM Interaction}\label{sec:CSM}

Flaring may occur due to the interaction of the outgoing jet with shells of material from historical eruptions of the progenitor star towards the end of its life (e.g. through pulsational pair instability ejections \cite{Ofek14,Aamer24}), potentially consistent with the ultra-long GRB progenitors proposed by \cite{Marchant20}. If the observed flares can be produced in this way, the need for late central engine activity may be removed, dramatically changing the requirements placed on the central engine.

While flares from jet-CSM interactions may be expected (e.g. \cite{Mesler12}), accounting for radial structure may lower the density to the point where no flare is produced at all \cite{Gat13}. Flares produced by sudden changes in environment density are expected to have durations comparable to their epoch (i.e. $\Delta t / t > 1$; \cite{Nakar03}). In contrast, the late flares in GRB\,220706A are extremely narrow. Taking $\Delta t$ as the time between the \textit{Chandra} detection and the final XRT non-detection (1.5\,Ms) and $t$ to be the time of the final XRT detection (4.4\,Ms), $\Delta t / t \approx 0.34$, inconsistent with expectations for a forward shock that encounters a sudden change in environment density.

One process that has been predicted to produce $0.01 \leq \Delta t / t \leq 0.5$ is the interaction between a late jet from the central engine and the cocoon that was produced when the main jet propagated through the progenitor star \cite{Shen10}. However, this requires the central engine to be active around the time of the observed flare, so does not reduce the required engine activity timescale.

\subsubsection*{A Tidal Disruption Event}\label{sec:TDE}

One final possibility is that GRB\,220706A was powered by a star entering the tidal radius of a black hole and being terminally disrupted; a tidal disruption event (TDE). GRB\,220706A is spatially consistent with the nucleus of its host galaxy (Figure~\ref{fig:opticalfield}), so the disruption of a star by the central supermassive black hole may be a plausible power source. A small number of TDEs have been seen to drive relativistic jets and produce bright high-energy emission (e.g. \cite{Levan11,Cenko12,Brown15,Andreoni22,Pasham23}) with optical counterparts that display similar peak magnitudes and evolution timescales to our candidate SN (e.g. \cite{Levan16}). The recent GRB\,250702B \cite{Carney25,Gompertz25,Levan25,Neights25,OConnor25} has re-opened the progenitor debate for ultra-long GRBs, with many studies suggesting a TDE involving a star and a stellar- or intermediate-mass black hole (e.g. \cite{Beniamini25,Eyles-Ferris25,Granot26}). Such a system would not be confined to the nucleus of the galaxy, though the galactic environment in the case of GRB\,250702B was unprecedented among GRB host galaxies \cite{Gompertz25}. Compact objects merging with stars have been suggested to be a common progenitor between ultra-long GRBs and luminous FBOTs \cite{Villar26}.

Nevertheless, the similarity of GRB\,220706A and its associated SN to GRB 111209A/SN2011kl, for which the SN was confirmed spectroscopically \cite{Greiner15}, provides support for the stellar collapse interpretation. The X-rays seen alongside GRB\,220706A are fully consistent with the GRB population and not the jetted TDE population (Figure~\ref{fig:Lum_xray}). The optical counterpart is also very unlike the well-sampled one observed alongside the jetted TDE AT2022cmc \cite{Andreoni22,Pasham23}.

\section{Methods}\label{sec:methods}

\subsection{High-energy data}

\subsubsection{\textit{Swift}}

\emph{Swift} data were downloaded from the UKSSDC \cite{Evans07,Evans09}. The BAT data were processed using the standard reduction pipeline {\sc batgrbproduct}. BAT spectra are created using the \emph{Swift}-dedicated {\sc batbinevt} routine, which is part of NASA's High Energy Astrophysics Software (HEASoft; \cite{HEASoft}). We also apply the required systematic error vector and updated keywords, and create a response matrix for each spectrum using {\sc batdrmgen}. Additional XRT data were obtained under ToO programmes 17696, 17777 and 17823. This corresponds to observation IDs 01114937013 -- 01114937026, inclusive.

The automated XRT analysis shows clear detections in late epochs at higher count rates than the preceding epochs. We investigated these data points manually to confirm that the apparent flaring is real and is not an instrumental artifact. Rebinning the data to one bin per \emph{Swift} observation revealed that all of the detected X-rays were concentrated in two observations: 01114937013 and 01114937014. In the first of these, the image shows a clear point-like source, bright enough that the blind source-detection algorithm of LSXPS \cite{Evans23} detected it and determined a count-rate consistent with that in the GRB light curve. There are no hot pixels, contaminating sources or other causes for concern in the data. In observation 01114937014 the source was not found by the LSXPS blind search\footnote{Note that a blind non-detection does not mean the source is not present; blind searches are less sensitive than searches for known sources, see \cite{Evans23}.}, despite its higher flux. This is not unexpected: the exposure time of this observation was 66\%\ shorter than the previous one, and the background level is 25\%\ higher. While the detected photons are not strongly point-like, this is not uncommon when there are only few events, and there are no signs of hot pixels or otherwise artificially inflated backgrounds, thus we find no reason to doubt this datapoint.

The automated light curve binning is designed to achieve a minimum number of counts per bin, rather than the optimal S/N per bin \cite{Evans07},
thus it combines 01114937014 with later observations to try to accumulate more counts. We have rebinned the data to keep observations
01114937013 and 01114937014 as distinct bins, and then created the final upper limit by combining all of the remaining observations.

\subsubsection{NICER}

NICER observed GRB\,220706A between 2022 July 07 and 2022 August 16 for a total raw exposure of $\approx$30 ks distributed over seven ObsIDs. We began the analysis by downloading Level 1 data from the publicly available HEASARC archive. The data were then processed using the standard HEASoft NICERDAS pipeline. Raw data were reprocessed with \texttt{nicerl2}, applying all standard screening criteria, except for adopting more conservative thresholds on undershoot and overshoot rates (0--200 and 0--2.0, respectively). These quantities trace distinct detector processes---overshoots primarily reflect charged-particle interactions, while undershoots are dominated by detector resets driven by optical loading and other charge injection mechanisms. The stricter cuts were motivated by the relatively low flux of GRB~220706A ($\sim10^{-11}$--$10^{-12}$ erg\,cm$^{-2}$\,s$^{-1}$). The output of this stage consists of calibrated and screened event files (separately for night and day intervals), along with the associated filter (MKF) files required for downstream analysis. In this work we only used data from the night, which is less susceptible to background flares.

Time-resolved analysis was performed by first constructing light curves using the \texttt{nicerl3-lc} pipeline and applying Bayesian blocks algorithm \cite{blocks2} to identify statistically significant changes in count rate. The light curves were binned at 512~s resolution prior to segmentation, and the resulting block boundaries were used to define physically motivated time intervals for spectral extraction. For each time bin, all contributing ObsIDs were identified and merged using \texttt{niobsmerge}, combining cleaned event files, unfiltered event files, and housekeeping data into a single dataset per interval. The Good Time Intervals (GTIs) of the merged event files were then modified to exactly match the boundaries of each Bayesian block, ensuring that spectral products correspond strictly to the desired temporal segments.

Spectra and associated response files were extracted for each time interval using the \texttt{nicerl3-spect} pipeline, adopting the SCORPEON background modeling framework. This task produces source spectra, redistribution matrices (RMFs), effective area files (ARFs), and a parameterized background model in a self-consistent manner. In contrast to traditional background subtraction methods, SCORPEON implements a forward-modeling approach in which the background is represented by a physically motivated model that is fit simultaneously with the source emission. This model is trained primarily on environmental tracers such as COR\_SAX and overshoot rates and includes both instrumental and astrophysical background components \footnote{\url{https://heasarc.gsfc.nasa.gov/docs/nicer/analysis_threads/scorpeon-overview/}}. Such an approach is particularly important for NICER’s non-imaging detector, where the background is both time-variable and inseparable from the source signal.

\subsubsection{\emph{Chandra}}

We initiated follow-up observations with the \textit{Chandra} X-ray Observatory under Director's Discretionary Time (Observation ID: 27263). Observations consisted of a single 15~ks exposure with ACIS-S, starting at UTC 12:42:21 on 2022 August 16, 40.9 days after trigger. The X-ray afterglow was detected at $\mbox{RA(J2000)} = \mbox{00:02:25.1}$, $\mbox{Dec(J2000)} = -\mbox{16:24:38.5}$ with a 90\% localisation uncertainty of $0.8''$. The unabsorbed 0.3 -- 10\,keV flux at this epoch is $3.09^{+0.93}_{-1.16} \times 10^{-14}$\,erg\,cm$^{-2}$\,s$^{-1}$ (see Table~\ref{tab:opt_phot}), calculated using the {\tt srcflux} routine in {\tt ciao} v4.15 \cite{Fruscione06} and converted from the native 0.5 -- 7\,keV bandpass of ACIS for comparison with the XRT data.

\subsection{Optical and infra-red data}

\subsubsection{NOT}

We started observing the field of GRB\,220706A using the Nordic Optical Telescope (NOT), equipped with the ALFOSC camera, as soon as the field was visible from the Canary Islands, that is about 0.52 days after the GRB on 2022-07-07 \cite{2022GCN.32338....1D}. Data reduction was carried out using standard procedures. Only the object marked ``B'' in Fig.~\ref{fig:opticalfield} was visible -- this was later discarded as the GRB counterpart.

A series of further images were secured in the following weeks, looking for the emergence of a possible SN (Table~\ref{tab:opt_phot}), and a late-time template image was obtained in 2024 July (about 738 days after the GRB). Photometry was calibrated against the Pan-STARRS1 catalog \cite{2024TNSTR3317....1C}.

\subsubsection{GROND}

We performed follow-up observations with GROND starting on 2022 July 07 at 09:28:00 UT, 0.72 days after the GRB trigger. Data were reduced with a dedicated pipeline, as described in \cite{Kruhler08}. $g'r'i'z'$ observations were calibrated against the Pan-STARRS1 catalogue \cite{2024TNSTR3317....1C} and $JHK$ data against 2MASS stars \cite{Skrutskie06}. The optical/nIR counterpart was not detected. Observations are tabulated in Table~\ref{tab:opt_phot}.

\subsubsection{VLT}\label{subsec:VLT}

We first acquired three epochs of $R$-band observations with the ESO Very Large Telescope (VLT) FORS2 instrument between 2022-08-09 and 2022-09-24, approximately 33.7, 45.6, and 79.6 days post-trigger. Each epoch comprised of $3\times 300$ s in imaging mode. The data were processed using the standard \textsc{esoreflex} workflow for the FORS2 instrument \cite{2013A&A...559A..96F}. Photometric measurements were performed using the Astropy Photutils package \cite{2019zndo...2533376B}, and the photometry was calibrated against the Pan-STARRS1 catalog \cite{2024TNSTR3317....1C}. While there is a mismatch between the FORS2 and Pan-STARRS filters, colour terms were found to have little effect.

For image subtraction, effective point spread function (PSF) models were constructed using \textsc{EPSFBuilder} from Photutils, wherein stars were extracted from the images to generate an over-sampled PSF model, optimizing the accuracy of subsequent image subtraction. The PyZOGY algorithm \cite{2017zndo...1043973G} was employed, using late-time $R$-band imaging acquired on 2024 September 05 (04:44:43 UT) as a subtraction template. The template consists of a total exposure time of 1800~s ($6\times300$~s) and was obtained once the transient had completely faded ($\sim700$ days after the burst).

\subsubsection{GTC}

We obtained imaging with the 10.4-meter Gran Telescopio Canarias (GTC) in La Palma (Canary Islands, Spain). Optical imaging in the $g'$, $r'$, $i'$, $z'$ filters was performed with OSIRIS on 2022-08-10 and 2022-08-29, 34.5 and 53.4 days after the burst, respectively. Near infra-red imaging in the $J$ filter was performed with EMIR on 2022-07-08, 1.51 days after trigger. Deep, filter-matched late-time imaging at 1252.2 days post-burst, when the transient had faded, was used as a host-galaxy template for image subtraction. Image subtraction for the GTC data was carried out using the same routine described in Section~\ref{subsec:VLT}, based on PSF-matched subtraction with the EPSBuilder and PyZOGY \cite{2017zndo...1043973G}. The latest GTC epoch (2025 December 08) was adopted as the template image. The results are summarised in Table~\ref{tab:opt_phot}.

Spectroscopic observations covering galaxies A and B were taken on 2022-08-09, 53.46 days after the GRB trigger, and consisted of $4\times1800$ s using grism R2500I and a slit of $0.8^{\prime\prime}$, resulting in a spectral coverage between 7335 and 10100 {\AA} at a resolving power of $\lambda/\delta\lambda\sim1900$. Data reduction was performed using standard procedures with self-developed pipelines. Data are corrected for bias, flat-field response, we then use HgAr, Ne and Xe lamps to perform a 2D wavelength correction and finally we flux calibrate using a spectrum of the spectrophotometric standard star Ross 640 obtained with the same spectral setup and a broad slit of 2.52$^{\prime\prime}$ during the same night.

\subsection{Radio and sub-mm data}

\subsubsection{NOEMA}

NOEMA observations were performed using a 10-antenna compact configuration (10D) using Bands 1 and 2. The first observation, in band 1 included two sidebands tuned at 74.256 GHz and 89.743 GHz, respectively and was performed 03:52 and 4:40 UT on 2022-07-08 (mean epoch 1.51 days after the burst). Following this, we observed in band 2, tuned at 136.257 GHz and 151.744 GHz between 6:19 and 7:43 UT (mean epoch 1.62 days after the burst). The primary flux calibrator was the carbon star MWC349. The bandpass calibrator was the strong quasar 3C454.3. Data reduction was done with the GILDAS programs CLIC and MAPPING. The side bands have r.m.s. noises of 52, 37, 67, and 184 $\mu Jy$, respectively. 

Although none of the side bands shows a signal that would allow us to identify a counterpart by itself (we would usually choose 5-$\sigma$ detections for this), the deepest image, at 89.743 GHz, has a peak point flux of $97\pm26$ ($3.7\sigma$ detection) consistent with the XRT position and with galaxy A (see the green ellipse in Figure~\ref{fig:opticalfield}). Furthermore, by fixing the NOEMA measurement to the VLA coordinates, the measurement remains a $3.7\sigma$ detection (at $86\pm26$). We consequently consider it a detection of the afterglow. Observations are recorded in Table~\ref{tab:opt_phot}.

\subsubsection{VLA}

VLA observations were performed on 2023-10-05, 455.7 days after the GRB trigger. Data were reduced using the Common Astronomy Software Applications Pipeline (\texttt{CASA}; \cite{2007ASPC..376..127M}). We used 3C147 for flux and bandpass calibration and J2357-1125 for complex gain calibration. Due to artifacts produced by a $\sim 20$mJy source $\sim 1.5$' from the afterglow position, we self-calibrated the target field with the prototype automatic self-calibration pipeline developed by the NRAO Science Ready Data Products Initiative\footnote{https://github.com/psheehan/auto\_selfcal}. We detect a clear source consistent with the afterglow location, and use the \texttt{pwkit/imtool} program \cite{2017ascl.soft04001W} to measure the flux density of the source, which we find to be $F_{\nu} = 53.2 \pm 12.8~\mu$Jy at a position of $\mbox{RA(J2000)} = \mbox{00:02:25.053}$, $\mbox{Dec(J2000)} = -\mbox{16:24:38.55}$ with a positional uncertainty of $0.18''$ ($1 \sigma$).

\subsection{Measuring $t_{\rm burst}$}\label{sec:tburst}

We measure $t_{\rm burst}$ following the method of \cite{Zhang14}. We use the combined BAT and XRT spectrally non-evolving light curves from the UKSSDC, binned to a signal-to-noise ratio (SNR) of 5 and extrapolated to the $0.3$ -- $10$\,keV bandpass of XRT \cite{Evans10}. Data were fit with the {\sc py-earth}\footnote{\href{http://contrib.scikit-learn.org/py-earth/index.html}{http://contrib.scikit-learn.org/py-earth/index.html}} implementation of the Multivariate Adaptive Regression Spline (MARS;\cite{Friedman91}). We find 549 GRBs with at least one power-law segment of $t^{-3}$ or steeper and a minimum of 6 data points in their light curves.

To place GRB\,220706A into context, we collect a sample of ultra-long GRBs. In the absence of a consensus definition of the class, we employ a simple cut in duration of $t_{\rm burst} > 10^4$\,s. The processing of {\em Swift} data has been shown to have a significant effect on the resultant light curve phenomenology \cite{Meredith23}. To mitigate this, where both are available, we require a measurement of $t_{\rm burst} > 10^4$\,s from the combined
BAT and XRT dataset \cite{Evans07} to be included. 20 GRBs pass this cut, and the 14 with a reported redshift are presented in Supplementary Table~\ref{tab:ULGRBs}. They include all of the `gold' sample ultra-long GRBs from \cite{Gendre19} as well as the archetypal ultra-long GRBs discussed in the literature \cite{Gendre13,Stratta13,Evans14}. Only one of these GRBs has X-ray coverage as late as was obtained for GRB\,220706A in the rest frame: the much lower redshift GRB 130925A ($z = 0.348$). We note the presence of possible flaring behavior at a similar epoch and luminosity to the late flares of GRB\,220706A. However, these are low significance fluctuations above the bright afterglow.

\subsection{Time-resolved high-energy spectroscopy}\label{sec:spectra}

To investigate the spectral evolution of GRB\,220706A, we divide the high energy observations from \emph{Swift} into discrete time bins. These are initially 2\,s wide, but are broadened as the GRB fades to maintain sufficient signal-to-noise in the bins. Each spectrum is fit in {\tt xspec} with either a power-law model (\texttt{pow}) with $\chi^2$ statistics for $\gamma$-rays or an absorbed power-law model (\texttt{tbabs * ztbabs * pow}) with Cash statistics \cite{Cash79} for X-rays. Milky Way absorption was fixed to $2.4\times 10^{20}$\,cm$^{-2}$ \cite{Willingale13} and abundances were taken from \cite{Wilms00}. We also fit an absorbed Band function \cite{Band93}, but find this only provides a statistical improvement of more than $3\sigma$ (i.e. a probability of chance improvement of $p < 0.003$) over the power-law fit in one spectral bin. The full suite of best spectral fits are shown in Extended Data Table~\ref{tab:spec_series} and the top panel of Figure~\ref{fig:xrays}.

\subsubsection{Inferences on the environment and synchrotron cooling}\label{sec:inferences}

For GRB afterglows, it is typically assumed (e.g. \cite{Sari98}) that electrons are accelerated into a power-law distribution of energies with an index of $p$. The measured photon index in X-rays is then $\Gamma = (p+1)/2$ if the synchrotron cooling frequency ($\nu_c$) is above the X-ray frequency ($\nu_x < \nu_c$), or $\Gamma = p/2 + 1$ for $\nu_c < \nu_x$. Since we expect $2 \lesssim p \lesssim 3$, we can infer $\nu_c < \nu_x$ for the afterglow of GRB\,220706A because the photon index is greater than 2, which would lead to $p > 3$ for $\nu_x < \nu_c$. In the $\nu_c < \nu_x$ spectral regime, the expected temporal index is $\alpha = (3p + 2)/4$ regardless of the radial profile of the surrounding material (typically assumed to be either ISM-like with $\rho \propto r^0$, or wind-like with $\rho \propto r^{-2}$). The measured slope of $\alpha = 0.98 \pm 0.07$ suggests $p = 1.97 \pm 0.09$, on the lower end of the expected range.

The spectrum between $3 \times 10^4$ and $10^5$~s after trigger is found to be unusually soft, with $\Gamma = 3.21^{+0.46}_{-0.40}$. This is inconsistent with a GRB afterglow, since it implies $p = 4.42^{+0.92}_{-0.80}$. This can be reconciled by using a Band function fit where the spectral peak is $\sim 1$\,keV, but the use of a more complicated model is not supported statistically. However, should this feature be real, the passage of $\nu_c$ through the XRT bandpass is the most likely cause, and would imply $\nu_x < \nu_c$ for the afterglow during the main flaring episodes from $10^2$ -- $10^4$s. This implies a shallower afterglow slope at this epoch, and would indicate an ISM-like environment because $\nu_c$ should evolve to higher energies with time in a wind-like medium. The lower value of $\Gamma$ measured at a few hundred seconds may support this interpretation. We further note that a broken power-law fit to the light curve between $200$ -- $10^6$~s with flares excluded is consistent with this possibility; a temporal slope of $\alpha_1 \approx 0.82$ is found, with a break at $\sim 2.3 \times 10^4$~s and $\alpha_2 \approx 1.06$. The change in $\alpha$ of $0.24$ is consistent with the theoretically expected $\alpha_2 - \alpha_1 = 0.25$. However, these parameters are poorly constrained by the fit. Similar spectral softening was observed in the ultra-long GRB\,130925A (e.g. \cite{Evans14,Piro14}), although GRB\,130925A remained spectrally soft whereas GRB\,220706A returns to more canonical values for the photon index from $\sim 10^5$\,s.

\subsubsection{Time-resolved analysis of the bright XRT flare}\label{sec:XRTflare}

To investigate the engine powering the bright flare between 4000 to 5000s after trigger, the WT mode data are divided into six 50-s bins between 4350 and 4650~s and fit with power-law (PL), cutoff power-law (CPL), Band function \cite{Band93} and PL + blackbody (PL+BB) models. The intrinsic absorption column, $N_{H,i}$, is tied between all epochs, since it is not expected to evolve on timescales this short. The best fits are shown in Supplementary Table~\ref{tab:time_resolved}.

We use Akaike's Information Criterion (AIC;] \cite{Akaike74}) to measure the relative support for each model from the data. This is expressed as $\Delta_{\rm AIC} = {\rm AIC}_{\rm model} - {\rm AIC}_{\rm min}$, where the best supported model always has $\Delta_{\rm AIC} = 0$. Models with $\Delta_{\rm AIC} > 4$ have little evidence to support them, and those with $\Delta_{\rm AIC} > 10$ are disfavored \cite{Burnham04}. Notably, the PL model is disfavored at all epochs. Its value evolves from $\Gamma = 1.78 \pm 0.05$ in the first epoch to $\Gamma = 2.02 \pm 0.04$, indicating the presence of the $\nu F_{\nu}$ spectral peak in the XRT bandpass, and hence the need for a more complicated model. The Band function is also rejected in the first epoch. While this could be interpreted as the peak not yet having entered the band, a break in the spectrum is in fact preferred; the model with the lowest AIC is the CPL, with a cutoff energy of $E_{\rm cut} = 2.86^{+0.67}_{-0.46}$\,keV. Indeed, across the six spectral epochs, the CPL model is the best supported, with a maximum $\Delta_{\rm AIC}$ of $2.6$. The PL+BB model is similarly well supported, though it is mildly disfavored in epoch 2, with $\Delta_{\rm AIC} = 6.3$. However, in its favor are photon indices consistent with expectations from synchrotron theory for the $\nu_m < \nu_x < \nu_c$ spectral regime. The values of $\Gamma$ in the CPL model are much more scattered, fluctuating around $\Gamma \sim 1$.

\subsubsection{Time-resolved analysis of the NICER flare}\label{sec:NICERflare}

Similar to the \textit{Swift}/XRT spectra, each NICER spectrum (0.25--12~keV) was modeled in \texttt{XSPEC} using the \texttt{ztbabs*tbabs*pow} model, with the Galactic absorption fixed at $2.4\times10^{20}$~cm$^{-2}$. Following the recommendations of the NICER data analysis threads, we adopted the PG-statistic for spectral fitting rather than $\chi^{2}$, as it provides a more appropriate likelihood framework for low-count data. In several time intervals, we observed a clear excess confined to a single spectral bin in the 0.5--0.65~keV range. This feature is a known systematic effect associated with foreground oxygen/neon emission from the Earth's atmosphere. At present, it is not possible to determine \emph{a priori} which observations are affected. The SCORPEON background modeling framework accounts for this feature through an additional spectral component with a free normalization. We therefore allowed this parameter to vary freely in all time-resolved fits. The results of this analysis are summarized in Supplementary Table~\ref{tab:NICER}.

In Epoch 2, we identify a systematic Gaussian-like residual centered at $\sim$2.5~keV (observer frame). We are not aware of any previously reported instrumental or background features at this energy in NICER data. However, given the non-imaging nature of NICER, the background cannot be independently measured, and we cannot definitively establish whether this feature is intrinsic to the source or arises from an unmodeled background component. Adding this Gaussian component to the absorbed power-law model improves the PG-statistic by $\Delta$PG = 74 for three additional degrees of freedom, corresponding to a statistically significant improvement in terms of $\Delta$AIC. The inclusion of this component also impacts the best-fit continuum parameters: the absorbing column decreases to $<4\times10^{20}$~cm$^{-2}$, and the photon index softens to $0.93^{+1.52}_{-0.12}$. The Gaussian component itself is characterized by a rest-frame centroid energy of $5.3^{+0.3}_{-0.3}$~keV and a width of $1.5^{+4.2}_{-0.6}$~keV. See Supplementary Figure~\ref{fig:line}.

A small number of emission features in GRB spectra have been reported in the literature \cite{2000Sci...290..955P, 2000ApJ...545L..39A,Campana24,Ravasio24}, but these detections remain controversial, and the community has generally been skeptical given their rarity and marginal statistical significance (e.g. \cite{Hurkett08}). Outside the GRB context, emission features in the $\sim$4--7~keV range are commonly observed in active galactic nuclei and are often interpreted as signatures of localized hotspots in the inner accretion flow \cite{2015A&A...574A.117M, 2004MNRAS.355.1073I, 2007A&A...475..155G}. While the energy scale differs in GRBs, one possible interpretation is that the feature we observe is associated with a transient, localized region of enhanced emission during the flaring episode. We emphasize, however, that this interpretation is speculative and to our knowledge no comparable feature has been robustly detected in a GRB. We therefore report it here with appropriate caution.

\backmatter

\bmhead{Supplementary information}

Supplementary information is available for this manuscript. The supplementary information file contains Supplementary Table 1~--~3, Supplementary Figure 1, and their references.

\bmhead{Acknowledgements}

This work made use of data supplied by the UK Swift Science Data Centre (UKSSDC) at the University of Leicester.
This research has made use of data obtained from the Chandra Data Archive provided by the Chandra X-ray Center (CXC).
Based on observations made with the Gran Telescopio Canarias (GTC), installed at the Spanish Observatorio del Roque de los Muchachos of the Instituto de Astrofísica de Canarias, on the island of La Palma, under programmes GTCMULTIPLE2H-22A and GTCMULTIPLE4B-25B. This work is partly based on data obtained with the IRAM NOEMA interferometer under the project codes s22bf001 and s22bf002. IRAM is supported by INSU/CNRS (France), MPG (Germany) and IGN (Spain). Based on observations made with the Nordic Optical Telescope, owned in collaboration by the University of Turku and Aarhus University, and operated jointly by Aarhus University, the University of Turku and the University of Oslo, representing Denmark, Finland and Norway, the University of Iceland and Stockholm University at the Observatorio del Roque de los Muchachos, La Palma, Spain, of the Instituto de Astrofisica de Canarias. The National Radio Astronomy Observatory and Green Bank Observatory are facilities of the U.S. National Science Foundation operated under cooperative agreement by Associated Universities, Inc.

BPG and Dimple acknowledge support from STFC grant No. ST/Y002253/1. 
BPG and DO acknowledge support from The Leverhulme Trust grant RPG-2024-117. 
AdUP acknowledges support from the Programme National Astro of CNRS/INSU with INP and IN2P3, co-funded by CEA and CNES through the Thematic Actions "Phénomènes Extrêmes et Multi-messagers" (PEM), “Physique et Chimie du Milieu Interstellaire” (PCMI) and "Cosmologie et Galaxies" (CG) of INSU Programme National “Astro".
DBM and DW acknowledge support from the Danish National Research Foundation under grant DNRF140 and are co-funded by the European Union (ERC, HEAVYMETAL, 101071865). Views and opinions expressed are, however, those of the authors only and do not necessarily reflect those of the European Union or the European Research Council. Neither the European Union nor the granting authority can be held responsible for them. The Cosmic Dawn Center (DAWN) is funded by the Danish National Research Foundation under grant DNRF140.
ANG acknowledges logistic support by the Th\"uringer Landessternwarte
Tautenburg, Germany. Part of the funding for GROND (both hardware and
personnel) was generously granted by the Leibniz-Prize to G. Hasinger
(DFG grant HA 1850/28-1) and by the Th\"uringer Landessternwarte
Tautenburg.
AR acknowledges support by PRIN-MIUR 2017 (grant 20179ZF5KS).
RLCS acknowledges support from The Leverhulme Trust grant RPG-2023-240.
BR, IW and MEW are supported by the UKRI Science and Technology Facilities Council (STFC).
DRP was supported by NASA NICER grant 80NSSC25K0645. 
MN is supported by the European Research Council (ERC) under the European Union’s Horizon 2020 research and innovation programme (grant agreement No.~948381). PAE and KLP acknowledge support from the UK Space Agency.
GS is a Canadian SKA Scientist and is funded by the Government of Canada/est financé par le gouvernement du Canada.

\section*{Data availability}

\emph{Swift} data are publicly available from the UKSSDC. \textit{Chandra} data are publicly available from the CXC archives. The raw VLT, GTC, and NOT data are available from their respective archives\footnote{\url{http://archive.eso.org/cms.html}}\footnote{\url{https://gtc.sdc.cab.inta-csic.es/gtc/jsp/searchform.jsp}}\footnote{\url{https://www.not.iac.es/observing/forms/fitsarchive/}}. Reduced data and tools used in this study are available upon reasonable request to the authors.

\section*{Author contributions}

BPG led the study, is the PI of the \emph{Chandra}, XRT and VLA observations, performed the high energy analysis (excl. NICER), and led the interpretation and writing of the paper. NH led the optical and infra-red photometric analysis, identified the supernova, and contributed to the writing and interpretation. DRP is the PI of the NICER observations and led the modeling, analysis and writing of sections related to them. AdUP led the GTC observational campaign, performed the spectroscopic analysis of these data, and is the PI of the NOEMA observations. DBM coordinated the observational campaign, is Co-PI of the VLT data, and is the PI of the NOT data, for which he took the photometric measurements. PAE and KLP performed the detailed verification of the late XRT detections and contributed to the high energy analysis. BR contributed to the VLT photometric analysis. JFAF contributed to the reduction and handling of the GTC data. MB led the execution and processing of the NOEMA observations. AJL made key contributions to the scientific interpretation and contributed to the optical data analysis. GS led the execution and processing of the VLA observations. NRT is the Co-PI of the VLT observations, made key contributions to the scientific interpretation, and contributed to the supernova identification and modeling. CCT is the PI of the GTC observations. SK, AMNG and AR led the GROND observations. All authors contributed to the scientific interpretation and the development of the text.

\section*{Conflict of interest}

The authors declare no conflict of interest.

\begin{appendices}

\setcounter{table}{0}
\renewcommand{\thetable}{\arabic{table}}
\makeatletter
\renewcommand{\fnum@table}{\textbf{Extended Data Table~\thetable}}
\makeatother

\section{Extended Data}\label{secA1}

\begin{sidewaystable}
    \centering
    \begin{tabular}{ccccccccccc}
    \hline\hline
    $t_{\rm start}$ & $t_{\rm stop}$ & Origin & Model & N & $\Gamma$ & $N_{H,i}$ & fit & dof \\
    (s) & (s) & & & (photons\,keV$^{-1}$\,cm$^{-2}$\,s$^{-1}$) & & ($10^{22}$\,cm$^{-2}$) & statistic & \\
    \hline
    0 & 2 & BAT & PL & $6.95^{+6.61}_{-3.39}$ & $1.80^{+0.19}_{-0.18}$ & & 49.80 & 56 \\
    2 & 4 & BAT & PL & $3.81^{+3.69}_{-1.90}$ & $1.65^{+0.19}_{-0.18}$ & & 57.50 & 56 \\
    4 & 6 & BAT & PL & $2.59^{+2.43}_{-1.28}$ & $1.55^{+0.18}_{-0.18}$ & & 63.43 & 56 \\
    6 & 8 & BAT & PL & $2.92^{+5.35}_{-1.92}$ & $1.68^{+0.29}_{-0.28}$ & & 59.82 & 56 \\
    8 & 10 & BAT & PL & $3.59^{+5.86}_{-2.23}$ & $1.77^{+0.28}_{-0.26}$ & & 46.98 & 56 \\
    10 & 15 & BAT & PL & $7.07^{+9.04}_{-3.91}$ & $2.00^{+0.24}_{-0.22}$ & & 53.93 & 56 \\
    15 & 20 & BAT & PL & $3.38^{+6.12}_{-2.16}$ & $1.89^{+0.30}_{-0.28}$ & & 50.98 & 56 \\
    20 & 40 & BAT & PL & $3.78^{+11.9}_{-2.80}$ & $2.17^{+0.43}_{-0.38}$ & & 51.77 & 56 \\
    \hline
    100 & 150 & XRT/WT & PL & ($5.66^{+1.25}_{-0.99}$) $\times 10^{-2}$ & $1.90^{+0.20}_{-0.19}$ & $0.81^{+0.30}_{-0.24}$ & 138.30 & 169 \\
    150 & 200 & XRT/WT & PL & ($5.84^{+1.77}_{-1.28}$) $\times 10^{-2}$ & $2.25^{+0.27}_{-0.25}$ & $1.17^{+0.46}_{-0.38}$ & 92.28 & 140 \\
    200 & 500 & XRT/PC & PL & ($2.00^{+0.55}_{-0.41}$) $\times 10^{-2}$ & $1.99^{+0.23}_{-0.21}$ & $2.58^{+0.73}_{-0.62}$ & 100.34 & 140 \\
    500 & 750 & XRT/PC & PL & ($5.73^{+1.78}_{-1.29}$) $\times 10^{-2}$ & $1.82^{+0.24}_{-0.22}$ & $2.30^{+0.84}_{-0.71}$ & 118.34 & 134 \\
    750 & 1000 & XRT/PC & PL & ($4.54^{+1.06}_{-0.82}$) $\times 10^{-2}$ & $1.60^{+0.20}_{-0.19}$ & $1.41^{+0.51}_{-0.42}$ & 103.15 & 132 \\
    4000 & 5000 & XRT/WT & PL & $0.80 \pm 0.02$ & $1.88 \pm 0.02$ & $2.69 \pm 0.08$ & 749.33 & 773 \\
    $1 \times 10^4$ & $2 \times 10^4$ & XRT/WT & PL & ($1.12^{+0.11}_{-0.09}$) $\times 10^{-1}$ & $2.37^{+0.09}_{-0.09}$ & $1.65^{+0.16}_{-0.15}$ & 338.09 & 392 \\
    $1 \times 10^4$ & $2 \times 10^4$ & XRT/PC & PL & ($7.02^{+0.86}_{-0.75}$) $\times 10^{-2}$ & $2.39^{+0.12}_{-0.01}$ & $1.69^{+0.21}_{-0.19}$ & 271.01 & 290 \\
    $2 \times 10^4$ & $3 \times 10^4$ & XRT/PC & PL & ($1.52^{+0.55}_{-0.37}$) $\times 10^{-3}$ & $2.76^{+0.39}_{-0.35}$ & $1.70^{+0.68}_{-0.53}$ & 62.19 & 65 \\
    $3 \times 10^4$ & $1 \times 10^5$ & XRT/PC & PL & ($6.55^{+2.85}_{-1.81}$) $\times 10^{-4}$ & $3.21^{+0.46}_{-0.40}$ & $2.63^{+0.90}_{-0.73}$ & 67.82 & 73 \\
    $3 \times 10^4$ & $1.5 \times 10^5$ & NICER & PL & ($1.16^{+0.27}_{-0.12}$) $\times 10^{-4}$ & $1.92^{+0.29}_{-0.36}$ & $0.39^{+0.30}_{-0.24}$ & 116 & 141 \\
    $1.5 \times 10^5$ & $1 \times 10^6$ & XRT/PC & PL & ($4.88^{+2.31}_{-1.44}$) $\times 10^{-5}$ & $2.38^{+0.41}_{-0.36}$ & $2.41^{+1.11}_{-0.87}$ & 64.40 & 74 \\
    $3.52 \times 10^6$ & $3.54 \times 10^6$ & \emph{Chandra} & PL & ($2.19^{+5.20}_{-2.19}$) $\times 10^{-5}$ & $2.58^{+0.97}_{-0.84}$ & $9.73^{+7.21}_{-5.67}$ & 45.23 & 62 \\
    \hline\hline
    \end{tabular}
    \caption{Power-law spectral fits to the $\gamma$-ray and X-ray data. The redshift is fixed to $z = 0.8577$ and the Galactic absorption to $2.4 \times 10^{20}$\,cm$^{-2}$ \cite{Willingale13}. We use the abundances from \cite{Wilms00}. BAT data are fitted between 15 and 150\,keV, XRT and \textit{Chandra} between $0.3$ and 10\,keV, and NICER data between $0.25$ and 12\,keV.}
    \label{tab:spec_series}
\end{sidewaystable}

\clearpage

\section{Supplementary Information}\label{sec:SI}

\setcounter{table}{0}
\renewcommand{\thetable}{\arabic{table}}
\makeatletter
\renewcommand{\fnum@table}{\textbf{Supplementary Table~\thetable}}
\makeatother
\renewcommand{\figurename}{Supplementary Figure}

\begin{table}[h]
    \centering
    \begin{tabular}{cccc}
    \hline\hline
        GRB & log ($t_{\rm burst}$ / s) & $z$ & Redshift reference \\
    \hline
        050904 & 5.50 & $6.29$ & \cite{Kawai05} \\
        060218 & 4.10 & $0.03$ & \cite{Mirabal06}\\
        060607A & 5.23 & $3.082$ & \cite{Ledoux06} \\
        070110 & 4.61 & $2.352$ & \cite{Jaunsen07} \\
        071021 & 4.06 & $2.452$ & \cite{Kruhler12} \\
        090926A & 4.72 & $2.106$ & \cite{Malesani09} \\
        101225A & 5.08 & $0.847$ & \cite{Levan14} \\
        111209A & 4.68 & $0.677$ & \cite{Vreeswijk11} \\
        121027A & 4.36 & $1.773$ & \cite{Tanvir12} \\
        130925A & 4.45 & $0.348$ & \cite{Vreeswijk13} \\
        170714A & 4.36 & $0.793$ & \cite{deUgartePostigo17} \\
        200613A & 4.71 & $1.228$ & \cite{deUgartePostigo21} \\
        211024B & 4.39 & $1.1137$ & \cite{deUgartePostigo22_GCN} \\
        220706A & 4.75 & $0.8577$ & This work \\
    \hline\hline
    \end{tabular}
    \caption{A sample of ultra-long GRBs with a reported redshift, selected by a cut in $t_{\rm burst}$ \citep{Zhang14} of $10^4$\,s or greater when measuring with both XRT-only data \citep{Evans07} and the combined BAT+XRT data \citep{Evans10}.}
    \label{tab:ULGRBs}
\end{table}

\begin{sidewaystable}
    \centering
    \begin{tabular}{cccccccccc}
    \hline\hline
    $t_{\rm start}$ & $t_{\rm stop}$ & Model & N & $\Gamma$ & & & fit & dof & $\Delta_{\rm AIC}$ \\
    (s) & (s) & & (photons\,keV$^{-1}$\,cm$^{-2}$\,s$^{-1}$) & & & & statistic & & \\
    \hline
    4350 & 4400 & PL & $0.82 \pm 0.04$ & $1.78 \pm 0.05$ & & & 440.66 & 508 & 15.6 \\
    4400 & 4450 & PL & $0.82 \pm 0.03$ & $1.86 \pm 0.04$ & & & 533.96 & 584 & 6.2 \\
    4450 & 4500 & PL & $0.82 \pm 0.03$ & $1.80 \pm 0.04$ & & & 572.53 & 605 & 13.4 \\
    4500 & 4550 & PL & $0.82 \pm 0.03$ & $1.96 \pm 0.04$ & & & 506.69 & 575 & 16.1 \\
    4550 & 4600 & PL & $0.82 \pm 0.03$ & $1.94 \pm 0.04$ & & & 519.79 & 585 & 13.1 \\
    4600 & 4650 & PL & $0.82 \pm 0.03$ & $2.02 \pm 0.04$ & & & 499.17 & 560 & 11.7 \\
     & & & & \multicolumn{3}{c}{$N_{H,i} = (2.76 \pm 0.09$) $\times 10^{22}$\,cm$^{-2}$} \\
    \hline\hline
    $t_{\rm start}$ & $t_{\rm stop}$ & Model & N & $\Gamma$ & $E_{\rm cut}$ & & fit & dof & $\Delta_{\rm AIC}$ \\
    (s) & (s) & & (photons\,keV$^{-1}$\,cm$^{-2}$\,s$^{-1}$) & & (keV) & & statistic & & \\
    \hline
    4350 & 4400 & CPL & $0.80^{+0.05}_{-0.05}$ & $0.72^{+0.18}_{-0.18}$ & $2.86^{+0.67}_{-0.46}$ & & 423.04 & 507 & 0 \\
    4400 & 4450 & CPL & $0.74^{+0.04}_{-0.03}$ & $1.10^{+0.15}_{-0.15}$ & $4.36^{+1.29}_{-0.82}$ & & 525.80 & 583 & 0 \\
    4450 & 4500 & CPL & $0.78^{+0.04}_{-0.04}$ & $0.86^{+0.15}_{-0.15}$ & $3.29^{+0.68}_{-0.49}$ & & 559.34 & 604 & 2.3 \\
    4500 & 4550 & CPL & $0.75^{+0.04}_{-0.04}$ & $1.19^{+0.15}_{-0.15}$ & $4.23^{+1.27}_{-0.80}$ & & 490.49 & 574 & 1.9 \\
    4550 & 4600 & CPL & $0.71^{+0.03}_{-0.03}$ & $1.37^{+0.15}_{-0.15}$ & $6.41^{+3.28}_{-1.65}$ & & 507.25 & 584 & 2.6 \\
    4600 & 4650 & CPL & $0.80^{+0.04}_{-0.04}$ & $1.02^{+0.16}_{-0.16}$ & $2.97^{+0.65}_{-0.46}$ & & 487.37 & 559 & 1.9 \\
     & & & & \multicolumn{3}{c}{$N_{H,i} = (1.92 \pm 0.12$) $\times 10^{22}$\,cm$^{-2}$} \\
    \hline\hline
    $t_{\rm start}$ & $t_{\rm stop}$ & Model & N & $\Gamma$ & $E_c$ & $\beta$ & fit & dof & $\Delta_{\rm AIC}$ \\
    (s) & (s) & & (photons\,keV$^{-1}$\,cm$^{-2}$\,s$^{-1}$) & & (keV) & & statistic & & \\
    \hline
    4350 & 4400 & Band & ($0.70^{+1.01}_{-0.42}$) $\times 10^{-1}$ & $0.52^{+0.19}_{-0.19}$ & $2.46^{+0.38}_{-0.36}$ & $9$ & 423.04 & 506 & 22.0 \\
    4400 & 4450 & Band & ($1.24^{+1.37}_{-0.66}$) $\times 10^{-2}$ & $0.88^{+0.16}_{-0.16}$ & $3.42^{+0.79}_{-0.56}$ & $9$ & 528.57 & 582 & 4.8 \\
    4450 & 4500 & Band & $0.39^{+1.13}_{-0.30}$ & $0.19^{+0.29}_{-0.36}$ & $1.64^{+0.54}_{-0.40}$ & $2.11^{+0.16}_{-0.12}$ & 555.08 & 603 & 0 \\
    4500 & 4550 & Band & ($2.18^{+6.12}_{-1.55}$) $\times 10^{-2}$ & $0.78^{+0.27}_{-0.28}$ & $2.50^{+1.03}_{-0.64}$ & $2.27^{+0.37}_{-0.17}$ & 487.28 & 573 & 0.7 \\
    4550 & 4600 & Band & ($1.28^{+3.57}_{-0.96}$) $\times 10^{-2}$ & $0.89^{+0.24}_{-0.28}$ & $2.78^{+1.26}_{-0.79}$ & $2.05^{+0.13}_{-0.11}$ & 502.67 & 583 & 0 \\
    4600 & 4650 & Band & $0.21^{+1.45}_{-0.16}$ & $0.34^{+0.33}_{-0.41}$ & $1.51^{+0.49}_{-0.41}$ & $2.35^{+0.19}_{-0.15}$ & 483.50 & 558 & 0 \\
     & & & & \multicolumn{3}{c}{$N_{H,i} = (1.63^{+0.15}_{-0.13}$) $\times 10^{22}$\,cm$^{-2}$} \\
    \hline\hline
    $t_{\rm start}$ & $t_{\rm stop}$ & Model & N & $\Gamma$ & $kT$ & $R_{BB}$ & fit & dof & $\Delta_{\rm AIC}$ \\
    (s) & (s) & & (photons\,keV$^{-1}$\,cm$^{-2}$\,s$^{-1}$) & & (keV) & ($10^{12}$\,cm) & statistic & & \\
    \hline
    4350 & 4400 & PL+BB & $0.42^{+0.07}_{-0.07}$ & $1.66^{+0.15}_{-0.14}$ & $0.75^{+0.08}_{-0.07}$ & $1.08^{+0.72}_{-0.60}$ & 421.52 & 506 & 0.5 \\
    4400 & 4450 & PL+BB & $0.50^{+0.07}_{-0.07}$ & $1.70^{+0.12}_{-0.12}$ & $0.72^{+0.12}_{-0.10}$ & $0.87^{+0.77}_{-0.57}$ & 530.09 & 582 & 6.3 \\
    4450 & 4500 & PL+BB & $0.41^{+0.07}_{-0.07}$ & $1.54^{+0.11}_{-0.11}$ & $0.64^{+0.06}_{-0.05}$ & $1.31^{+0.84}_{-0.73}$ & 555.43 & 603 & 0.3 \\
    4500 & 4550 & PL+BB & $0.49^{+0.06}_{-0.06}$ & $1.81^{+0.11}_{-0.10}$ & $0.69^{+0.08}_{-0.07}$ & $0.97^{+0.71}_{-0.58}$ & 486.57 & 573 & 0 \\
    4550 & 4600 & PL+BB & $0.51^{+0.07}_{-0.07}$ & $1.73^{+0.09}_{-0.09}$ & $0.62^{+0.11}_{-0.08}$ & $0.98^{+0.89}_{-0.70}$ & 503.98 & 583 & 1.3 \\
    4600 & 4650 & PL+BB & $0.40^{+0.07}_{-0.07}$ & $1.75^{+0.11}_{-0.12}$ & $0.58^{+0.05}_{-0.05}$ & $1.46^{+1.00}_{-0.84}$ & 484.37 & 558 & 0.9 \\
     & & & & \multicolumn{3}{c}{$N_{H,i} = (1.90 \pm 0.15$) $\times 10^{22}$\,cm$^{-2}$} \\
    \hline\hline
    \end{tabular}
    \caption{A comparison of time-resolved spectral models fitted to the bright XRT flare at $\sim 4500$~s after trigger. The intrinsic $N_H$ is tied between epochs, with the best-fit value for each spectral model given in the table. The Galactic absorption column is fixed to $2.40 \times 10^{20}$\,cm$^{-2}$ \cite{Willingale13}. Abundances are taken from \cite{Wilms00}. The preferred model always has $\Delta_{\rm AIC} = 0$. Models with $\Delta_{\rm AIC} > 4$ are disfavored, while those with $\Delta_{\rm AIC} > 10$ are rejected by AIC \citep{Burnham04}. Accounting for all epochs, the CPL model is the best supported. The PL model is strongly rejected.}\label{tab:time_resolved}
\end{sidewaystable}

\begin{sidewaystable}
    \centering
    \begin{tabular}{ccccccccc}
    \hline\hline
    Epoch & t$_{\rm start}$ & t$_{\rm stop}$ & $\Gamma$ & N$_{H,i}$ & Flux & Luminosity & PGstat & dof \\
     & (s) & (s) & & ($10^{22}$\,cm$^{-2}$) & (erg\,cm$^{-2}$\,s$^{-1}$) & & & \\
    \hline
1 & 34121 &  96070 & 1.66$^{+0.31}_{-0.24}$ & 0.23$^{+0.12}_{-0.05}$ & (2.24$^{+0.52}_{-0.42}$) $\times 10^{-12}$ & (6.76$^{+0.65}_{-0.59}$) $\times 10^{45}$ & 119.6 & 110 \\
2 & 96070 &  107561 & 0.69$^{+0.05}_{-0.05}$ & <0.82 & (1.70$^{+0.12}_{-0.11}$) $\times 10^{-11}$ & (2.82$^{+0.20}_{-0.19}$) $\times 10^{46}$ & 215 & 107 \\
3 & 107561 &  126742 & 1.02$^{+0.11}_{-0.14}$ & <0.82 & (5.89$^{+0.87}_{-1.52}$) $\times 10^{-12}$ & (1.10$^{+0.08}_{-0.14}$) $\times 10^{46}$ & 147 & 111 \\
4 & 126742 &  132358 & 2.22$^{+0.83}_{-0.83}$ & <0.32 & (9.77$^{+0.72}_{-0.36}$) $\times 10^{-13}$ & (4.07$^{+0.29}_{-0.53}$) $\times 10^{45}$ & 107 & 112 \\
    \hline\hline
    \end{tabular}
    \caption{Time-resolved spectral fits of the NICER data, using a power-law model. The Galactic absorption is fixed to $2.4 \times 10^{20}$\,cm$^{-2}$ \citep{Willingale13}.}
    \label{tab:NICER}
\end{sidewaystable}

\begin{figure}[H]
    \centering
    \includegraphics[width=\columnwidth]{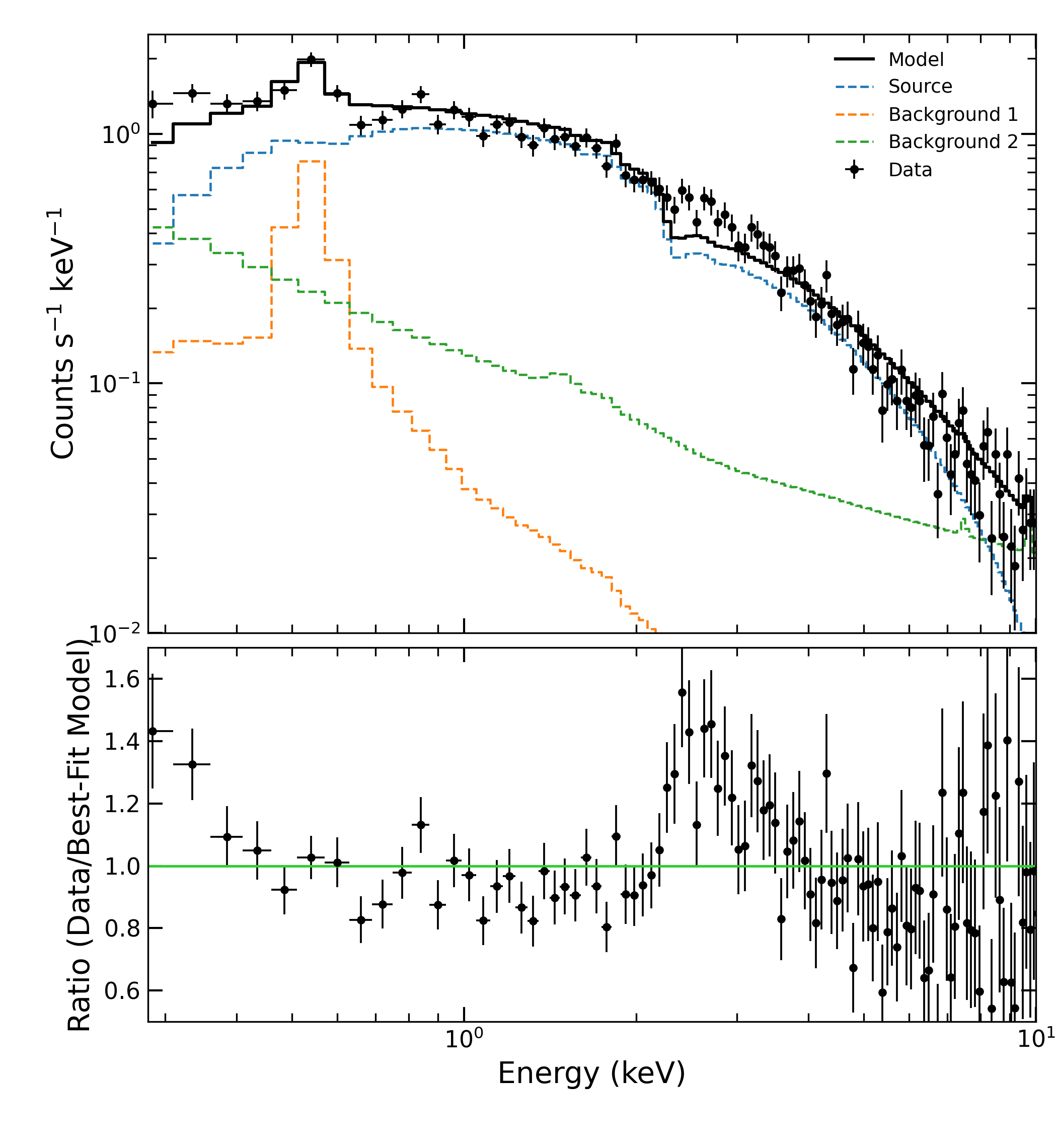}
    \caption{\textit{Top}: Observed X-ray spectrum (source + background; filled circles) of NICER Epoch 2, modeled with a power-law continuum (solid line) and two background components from the SCORPEON framework (dashed green and orange curves). \textit{Bottom}: Data-to-model ratio, showing a clear excess at $\sim$2.5 keV (observer frame), suggestive of a potential spectral feature.}
    \label{fig:line}
\end{figure}

\end{appendices}

\clearpage

\bibliography{sn-bibliography}

\end{document}